\documentclass[%
 reprint,
 amsmath,amssymb,
 aps,
]{revtex4-2}

\usepackage{graphicx}
\usepackage{dcolumn}
\usepackage{bm}
\usepackage{hyperref}

\usepackage[dvipsnames]{xcolor} 
\usepackage{amssymb}
\usepackage[none]{hyphenat}
\usepackage{amsmath}
\usepackage[left]{lineno}
\usepackage{marvosym}

\usepackage{enumerate}
\usepackage{makecell}
\usepackage{booktabs}
\usepackage[normalem]{ulem}
\usepackage{caption}
\usepackage{float}
\setcitestyle{super}

\usepackage{xcolor}

\begin{document}
\preprint{APS/123-QED}
\title{Percolation criticality of amorphous-amorphous transitions in compressed glasses}

\author{J. Perradin$^1$, S. Ispas$^1$, R. Paredes$^{2,3}$, A. Hasmy$^{1,4}$, and B. Hehlen$^1$}

\address{$^1$Laboratoire Charles Coulomb (L2C), CNRS - Universit\'e Montpellier, 34095 Montpellier, France}
\address{$^2$Departamento de F\'{i}sica  y Matemáticas, Universidad Iberoamericana, 01219 Ciudad de México, Mexico.}
\address{$^3$Centro de Física, Instituto Venezolano de Investigaciones Científicas, Apdo. 21827, 1020A Caracas, Venezuela}
\address{$^4$Departamento de F\'{i}sica, Universidad Sim\'on Bol\'{i}var, Valle de Sartenejas, Caracas, Venezuela}

\date{\today}

\begin{abstract}
The low-to-high-density transition in compressed silica glass is investigated using percolation theory. 
Large-scale molecular-dynamics simulations of SiO$_2$ glasses, with system sizes of up to 10$^6$ atoms and pressures ranging from 0 to 35 GPa, were carried out to investigate the emergence of structural motifs and their growth to system-spanning length scales under compression. 
On this basis, we introduced long-range descriptors that complement conventional local and medium-range structural measures.
The results reveal critical percolation transitions of SiO$_Z$-SiO$_Z$  clusters with increasing coordination number $Z$. 
The critical exponents of the lowest-pressure transitions, {\it i.e.}, tetrahedral and pentahedral
networks, are consistent with standard percolation predictions, whereas for higher-coordinated polyhedra the data collapse suggests possible deviations that are tentatively associated with a rigidity-percolation mechanism.
SiSi$_z$-SiSi$_z$ clusters were also analyzed using the non-bonded approach. 
Bonded and non-bonded approaches complement each other in a particularly illuminating way for describing pressure-induced structural transformations and common mechanisms shared by bonded glasses, such as SiO$_2$, and non-bonded glasses, such as amorphous ice.

\end{abstract}

\maketitle

\section{Introduction}
Elucidating the mechanisms underlying structural transformations in compressed glasses continues to pose a significant challenge across materials science, condensed-matter physics, and geophysics. 
Tetrahedral-based materials such as water, chalcogenides, metallic and oxide glasses exhibit complex pressure-induced structural changes that hinder our current understanding of amorphous-amorphous transformation~\cite{Loe2009,Mac2014, Lak2000}.
Unlike the sharp phase transitions between crystalline polymorphs, amorphous-amorphous transformations upon compression display a progressive evolution of local structural motifs, resulting in distinct glassy states known as polyamorphs.
This results in a gradual increase in the sample density and average coordination number $Z$, which represents the average number of nearest neighbors of a given atomic species.

At ambient pressure, the structure of vitreous silica (or amorphous silica, $a$-SiO$_2$)  consists of SiO$_4$ tetrahedra linked at their corners by bridging oxygen atoms. At the medium-range scale, the Si-O-Si connected network forms large Si-O-Si rings of various sizes, yielding an open structure that is prone to substantial densification during compression \cite{varshneya2019fundamentals}.
In addition, the scaffold is floppy due to the presence of rigid unit modes (RUMs)\cite{Tra2002}, which consist of low-frequency vibrations and translations of the nearly undeformed tetrahedra associated with the boson peak \cite{Buc1984,Heh2000,Sch2007}, and structural relaxation processes~\cite{Ruff2022}. 
This flexibility with respect to RUMs allows the structure  to compact easily under compression.
As pressure increases, the structure becomes rigid against RUMs, which are progressively suppressed\cite{Cou2003}. 
From then on, densification proceeds mainly through distortion of the local tetrahedral structure \cite{Tra2003}, concomitant with the onset of increasing coordination numbers for both silicon and oxygen.
These processes are likely essential for understanding the compressibility peak around 3\,GPa. 
The latter share striking similarities with compressed amorphous semiconductors~\cite{Der2021} and non-bonded glasses, i.e., amorphous ice, $a$-H$_2$O \cite{Ama2016}.
In both cases, the compressibility $\kappa_{_T}$ exhibits a peak close to the region where the density-pressure relation becomes steeper, indicating the onset of the low-to-high density transition~\cite{Has2025}.
However, the out-of-equilibrium nature and the complexity of polyamorphism in the glassy state of $a$-SiO$_2$ have so far prevented the identification of a scale-invariant physical quantity (i.e., an order parameter) for the disorder–disorder transformation, as is typically done for liquid and crystal phase transitions.

Recently, simulations of SiO$_2$ glasses have shown that under compression, a large cluster with increased polyhedricity emerges and percolates, indicating the onset of a new dominant ``phase"~\cite{Has2021}. 
Percolation provides an elegant and simple statistical-mechanical framework  for studying phase transitions in which criticality is associated with the formation of a system-spanning connected structure. It has been used to model connectivity-driven phenomena across numerous physical, biological, and environmental  systems~\cite{Sab2015}.
In amorphous materials, percolation transition has been observed in structural transformations of supercritical~\cite{Par2007,Ber2008}, supercooled and glassy water~\cite{Bro2003,Has2025}, while some structural features reminiscent of percolation transitions have also been reported for bonded metallic glasses~\cite{Che2015}. 
Despite progress in studying amorphous transitions within the percolation framework, the criticality of percolation in glasses remains unknown. 
Estimating critical exponents would clarify how connectivity, cluster sizes, and mechanical rigidity evolve near the threshold, revealing universal scaling. 
This helps classify disordered systems into universality classes, guiding models of network formation, rigidity percolation, and critical phenomena, and deepening understanding of how large-scale mechanical stability emerges from local interactions. 
A key question, therefore, is whether structural transitions in glasses are substance-dependent or if they fall within the same universality class as standard (random) uncorrelated percolation.
In the pioneering work of Hasmy {\it et al.}~\cite{Has2021}, the limited system size inherent to the {\it ab initio}-based approach prevented access to this information.
In this work, we employ classical molecular-dynamics simulations in large systems to investigate structural transformations in compressed SiO$_2$ glasses and amorphous ice, and estimate their percolation critical exponents. 
Structural descriptors based on the coordination numbers of SiO$_Z$ polyhedra have been used for $a$-SiO$_2$, and OO$_Z$ structures for $a$-H$_2$O.   
In addition, in $a$-SiO$_2$ we have also considered SiSi$_Z$ polyhedra as local descriptors in order to, first, complement the results obtained using the conventional SiO$_Z$-based analysis, and, second, to parallel the case of amorphous ice, in order to envisage the possibility of defining a unique structural framework to be used for all polyhedral systems.

\section{Methods}
\subsection{Computational details}

The molecular-dynamics simulations of SiO$_2$ glasses were performed using the SHIK pair potential~\cite{sundararaman2018new} implemented in the LAMMPS package~\cite{thompson2022lammps}.
A time step of 1.6~fs was used, and we considered, for the short-range and long-range terms, the same cutoffs as in previous works~\cite{sundararaman2018new, zhang2020critical}.
We considered seven system sizes containing 1008, 3024, 8064, 15\,120, 27\,216, 96\,000, and one million atoms, and five samples per size were generated. 
Starting from an initial random configuration, we first  equilibrated at 3500~K in the canonical ensemble (NVT) using a cubic box that corresponds to the experimental silica density  at room temperature, i.e., 2.2~g/cm$^3$~\cite{varshneya2019fundamentals}.
The length of this run was $560$~ps. We then quenched the sample to 3000~K with a quench rate of about $3$~K/ps. At 3000~K, we switched to the isothermal-isobaric ensemble (NPT) with a first short run of $\approx 80$~ps during which the pressure was reduced to zero, and then with an NPT run of 320~ps at 0~GPa.
The above-mentioned length runs were sufficient in order to reach a linear diffusive regime. 
Subsequently, the samples were quenched in the NPT ensemble at a rate of 1~K/ps, a cooling rate sufficiently slow to ensure that it does not significantly affect the resulting properties at zero pressure~\cite{zhang2020critical, shih2020}.
At room temperature, an NPT relaxation was performed for 160~ps to relieve some residual stress.
The average densities at the end of this multi-stage melt-and-quench procedure were in very good agreement with the experimental room temperature density, as already stated in previous works~\cite{sundararaman2018new}.
After an additional NVT run (about 80~ps), the resulting samples were quasi-statically compressed by gradually reducing their volume by 1\% in order to reach a pressure of about 35~GPa. After each compression, NVT runs at 300~K were carried out for 100~ps.

For modeling amorphous ice, we follow the same protocol used in previous studies to compress five samples consisting of 16\,384 H$_2$O molecules at 124~K~\cite{Gar2021,Has2025}.
The molecular-dynamics simulations were performed using the TIP4P force field~\cite{Tip4p} implemented in the Gromacs package~\cite{Gromacs}. This model gives a density of 0.9964~g/cm$^3$ under ambient conditions, which is very close to the experimental water density of 0.997~g/cm$^3$. 
The procedure involved equilibrating liquid water at 300~K at ambient pressure, followed by isobaric cooling to 124~K in steps of 1~K per nanosecond, and subsequent isothermal compression to 15~kbar in increments of 0.1~kbar per nanosecond. 
Data for compressed amorphous ice, with smaller sizes (336 to 8192 molecules), are taken from Ref.~\cite{Has2025}.

\subsection{Structural units and cluster analysis}

For amorphous silica, we first adopted the traditional way of describing the structure as a continuous network of SiO$_Z$ polyhedra with Si ions connected to $Z$ oxygen  ions with a high degree of covalency.
Within the percolation framework analysis, we call this approach ``the \textit{bonded model}''~\cite{Okeefe1981}. 
The coordination number $Z$ was computed using a cutoff distance of 2.3~\AA\ for the Si-O bond. A sensitivity test has shown that small variations in the Si–O cutoff have only a minor effect on the coordination numbers and percolation properties.
In the cluster analysis, we considered the following connectivities of these SiO$_Z$ polyhedra: corner-shared (CS), edge-shared (ES), and face-shared (FS).
In addition, we considered the case called hereafter ``$a$-stishovite'' corresponding to two SiO$_6$ octahedra sharing exactly two edges.
Alternatively, the structure of silica can be described as a network whose nodes are silicon atoms, with the local structures represented by SiSi$_Z$ polyhedra. 
The latter are defined using a cutoff distance of 3.5~\AA, which corresponds to the position of the first minimum in the Si–Si pair distribution function across the pressure range considered. 
This cutoff encompasses all first-neighbor Si--Si distances found in the crystalline counterparts quartz, cristobalite, and coesite-I. It is also consistent with coesite-IV, a high-pressure structure containing SiO$_4$, SiO$_5$, and SiO$_6$ units, which has been shown to provide a close crystalline analog of silica glass under moderate pressure~\cite{Has2021}.
In this structure, the first and second Si–Si coordination shells are broader but remain separated by a small gap between approximately 3.35 and 3.60\,\AA\, (range estimated  using the data reported in Ref.~\cite{Byk18}).
The choice of the cutoff in the non-bonded model of SiO$_2$ follows the same rationale as that used for $a$-H$_2$O~\cite{Has2025}.
In this context, we can consider the interactions between the Si atoms within the framework of a \textit{non-bonded} force model, rather than relying on \textit{bonded} (covalent) interactions~\cite{Okeefe1981}. 
It is also worth noticing that, by construction, SiSi$_Z$ structures are more distorted than SiO$_Z$ ones.   

Among the different schemes to describe the structure of water, some consider solely the oxygen-oxygen arrangements by defining ordered and disordered O-O tetrahedral structures \cite{Err2001,Wik2011}. 
This description is commonly used as it facilitates parallels between crystalline and amorphous ices~\cite{Pie2015,Monse2020,Mar2020}. 
Here, we propose an indicator based on the coordination number $Z$ of oxygen atoms, as done previously~\cite{Has2025}. 
This disregards the small, disordered hydrogen atoms and thus falls within the {\it non-bonded} model.
The structure of low-density water (LD) is composed of nearly perfect void-center oxygen tetrahedra of coordination number $Z_{\rm OO}=4$, while high-density (HD) structures consist of distorted configurations with $Z_{\rm OO}=5$--$7$.
The remaining units, with $Z\geq 8$, are denoted as very-high-density (VHD) structures. 
The O-O cutoff distance is fixed to 3.5~\AA,  
which lies in the flat minimum of the O–O pair distribution function between the first and second coordination shells and is widely used to identify LD and HD populations. This choice is discussed in detail in Ref.~\cite{Has2025}. No explicit sensitivity test on coordination numbers and percolation properties was performed for this cutoff in the present work. 
The cluster analysis also did not distinguish between the various possible connectivities of Z$_{OO}$ units.
The same definitions for LD, HD, and VHD structures are used for the SiSi$_Z$ cluster analysis of $a$-SiO$_2$.

To identify polyhedral clusters in both \textit{a}-H$_2$O and \textit{a}-SiO$_2$, and more generally for atomistic systems, we used a dedicated Python code, \textit{Nexus-CAT}, implementing a flexible union-find algorithm~\cite{perradin_nexus-cat_2026,perradin_polyamorphism_2025}.
The program identifies the nearest neighbors surrounding the central atom of interest---either silicon or oxygen---within a spherical region whose radius is set according to the cutoff values defined for each system and/or model.

The resulting list of neighbors, used to identify clusters, is filtered based on various criteria, specifically the coordination number $Z$ and eventually the number of shared oxygens needed to identify \textit{a}-stishovite clusters in \textit{v}-SiO$_2$.
Percolating clusters are detected using the method of Livraghi \textit{et al.}~\cite{livraghi_exact_2021}, as implemented in \textit{Nexus-CAT} (see documentation~\cite{perradin_nexus-cat_2026}).
With the resulting sets of clusters, the percolation properties such as the correlation length $\xi$, the order parameter $P_\infty$, the average cluster size $\langle S\rangle$, the mean largest cluster size $S_\text{max}$, etc., are calculated across sample sizes and averaged over the configurations of the system (see the Supplemental Material~\cite{supmat} and \textit{Nexus-CAT} documentation~\cite{perradin_nexus-cat_2026} for definitions of the percolation parameters). Unlike in previous works \cite{Has2021,Has2025}, the critical pressures were identified at the onset of cluster percolation across all three dimensions of the box.

Finite-size scaling methods are used to determine critical exponents and fractal dimensionality. 
Specifically, pyfssa~\cite{sorge_pyfssa_2015}, iminuit\cite{iminuit,james_minuit_1975}, and the Kawashima-Ito cost functions~\cite{kawashima_critical_1993} were used to determine critical parameters \textit{via} data-collapse methods.


\begin{figure}[ht]
    \includegraphics[width=\linewidth]{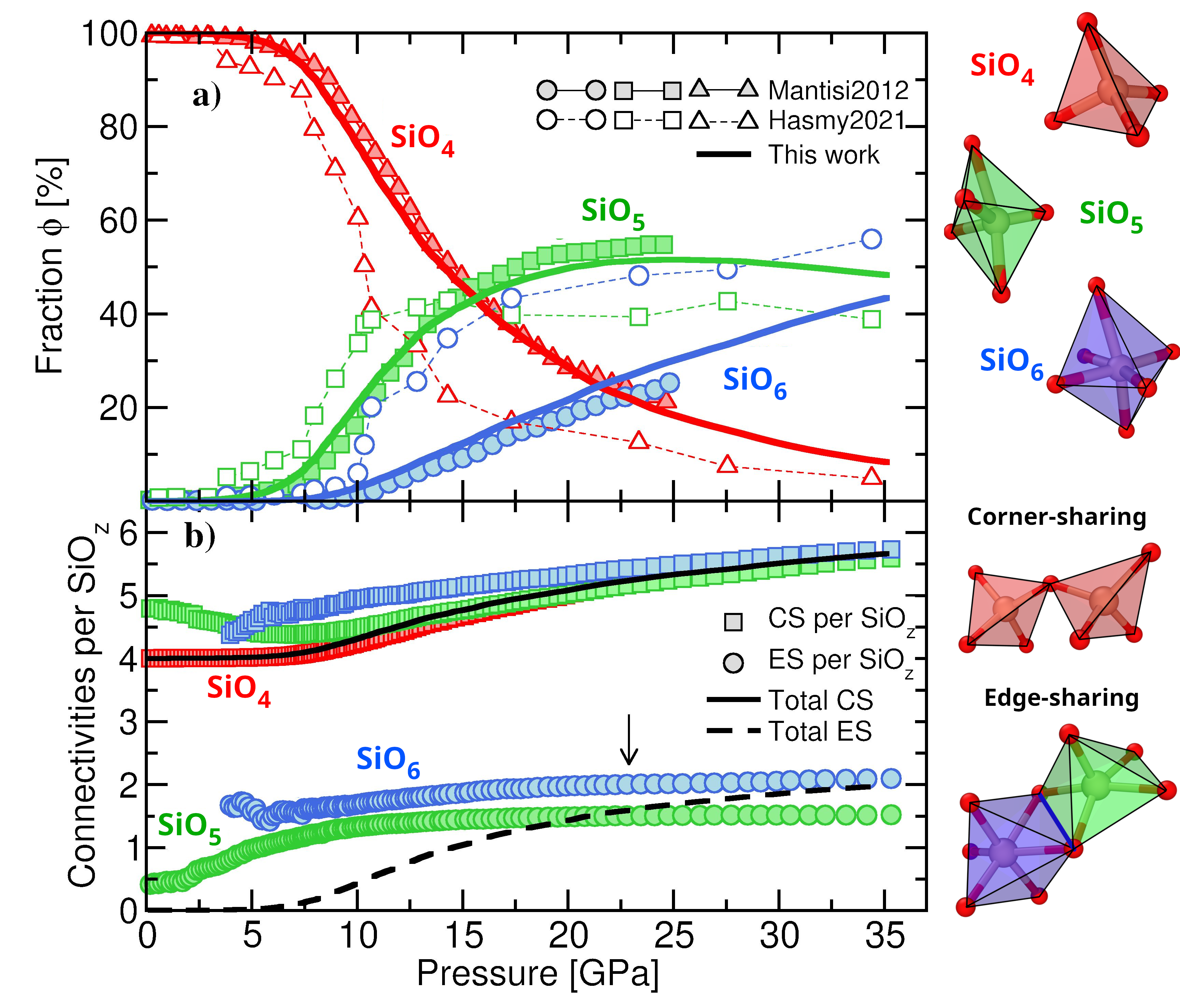}
    \caption{\textbf{(a)} Fraction $\phi$ of SiO$_Z$ polyhedra: SiO$_4$ (red), SiO$_5$ (green), and SiO$_6$ (blue). The data extracted from the present work (bold lines) are compared with those extracted from classical molecular-dynamics (MD) simulations (colored symbols and continuous lines) reported by Mantisi {\it et al.}~\cite{mantisi2012atomistic}  and DFTB~\cite{Has2021} (open symbols and dashed lines) MD studies.
    \textbf{(b)} Corner- and edge-sharing connectivities per SiO$_Z$. The arrow indicates the pressure at which the number of ES bonds equals two, as in crystalline stishovite \cite{Byk18}.}
    \label{fig:figure1}
\end{figure}
\section{Results}
This section first presents the pressure dependence of the local structural properties of SiO$_2$ and H$_2$O, with particular emphasis on coordination number and connectivity of Si units. The analysis is then extended to percolation, for which both bonded and non-bonded approaches are used for identifying the large-scale structural descriptors that govern the overall network organization.

\subsection{Local structures}

At ambient pressure, $a$-SiO$_2$ consists predominantly of a network of CS SiO$_4$ tetrahedra, with only a few residual SiO$_5$ polyhedra remaining from the liquid phase ($<$0.1\%).
The pressure dependence of the fraction $\phi$  of SiO$_Z$ polyhedra  is shown in Fig.~\ref{fig:figure1}a. 
The SHIK results (this work) are compared with the results obtained using a different force field~\cite{mantisi2012atomistic} and with quantum semi-empirical density functional tight binding (DFTB) calculations~\cite{Has2021}. 
Overall, the two classical pair potentials exhibit very similar behavior, whereas the DFTB method predicts a higher fraction $\phi$ of SiO$_6$ polyhedra at intermediate pressures, leading to a faster decrease in the number of SiO$_4$ tetrahedra.
Upon closer inspection, the fraction of SiO$_5$ pentahedra begins to increase around 3~GPa in DFTB, and at approximately 5~GPa in the two classical force fields, i.e., at a pressure just beyond the maximum of compressibility, observed experimentally at approximately 3~GPa~\cite{bridgman1938high}.
With increasing pressure, SiO$_6$ octahedra begin to emerge around 10 GPa in all cases, while some edge-sharing units appear during the early stage of SiO$_5$ and SiO$_6$ formation, as shown in Fig.~\ref{fig:figure1}b.  
However, the total number of ES units becomes significant only beyond 8 to 10~GPa, a pressure slightly higher than the onset pressure of the plastic regime, which occurs at 7 to 8~GPa in our models, compared with $\approx$10~GPa in experiments~\cite{Van2008,keryvin_constitutive_2014,rouxel_poissons_2008}.
Above 23~GPa, the number of ES bonds per SiO$_6$ evolves very smoothly with pressure and reaches the value of 2 [arrow in Fig.~\ref{fig:figure1}(b)], similar to crystalline stishovite.

\subsection{Percolation in SiO$_2$ glasses}

\begin{figure}[htbp]
   \begin{center}
       \includegraphics[width=\linewidth]{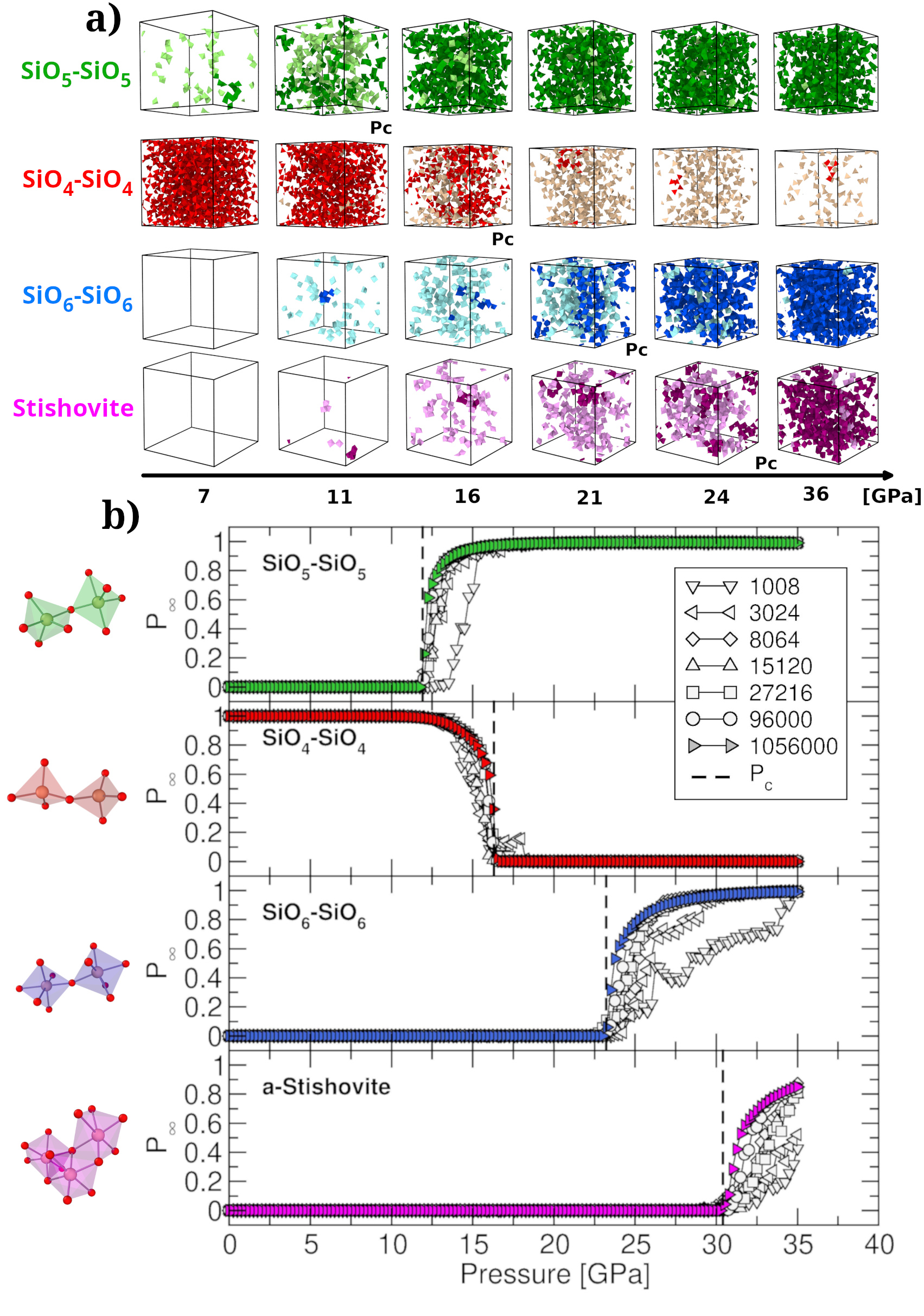}
   \end{center} 
    \caption{
     {\bf (a)} Snapshots of pressurization of an $a$-SiO$_2$ sample of 3024 atoms. Lighter colors label small SiO$_z$-SiO$_z$ clusters and darker colors correspond to the biggest cluster (green shades are SiO$_5$-SiO$_5$, red are SiO$_4$-SiO$_4$, blue are SiO$_6$-SiO$_6$, and purple are stishovite, i.e., SiO$_6$ connected by ES). {\bf (b)} Percolation order parameter $P_\infty$ of SiO$_z$-SiO$_z$ and $a$-stishovite clusters as a function of pressure for each simulation box size $L$. 
    The critical pressure $p_c$ (dashed lines) corresponds to the maximum of the correlation function $\xi(p)$ and the colored points to simulations performed for the largest system size.}
    \label{fig:figure2}
\end{figure}

Figure~\ref{fig:figure2}(a) shows various snapshots of the structural evolution upon compression of the SHIK glass within the bonded approach. 
Isolated SiO$_5$ pentahedra (green boxes) appear at low pressure and progressively replace  SiO$_4$ (red boxes) in the silica network. 
They form (SiO$_5$-SiO$_5$)$_n$ clusters and at a critical pressure $p_c \simeq$~12 GPa (Table~\ref{tab:table_pc}), the biggest one (dark green), called the \textit{spanning cluster}, percolates along the three dimensions of the box.
The (SiO$_4$-SiO$_4$)$_\infty$ and (SiO$_5$-SiO$_5$)$_\infty$ infinite clusters coexist up to  
$p_c\simeq$~16~GPa where the tetrahedral three-dimensional (3D) connected network depercolates [i.e., the (SiO$_4$-SiO$_4$)$_\infty$ cluster collapses] due to the increase in the fraction of SiO$_5$ pentahedra followed by that of SiO$_6$ octahedra (blue boxes).
On further increasing the pressure, a (SiO$_6$–SiO$_6$)$_n$ cluster percolates at $p_c \simeq 23$~GPa (dark blue), and at $p_c \simeq 30$~GPa an (SiO$_6$–SiO$_6$)$_\infty$ infinite cluster emerges, comprising two edge-sharing connections per octahedron, akin to crystalline stishovite (dark purple). 
This is indeed very close to the pressure at which the average number of ES per SiO$_6 \simeq 2$ [see Fig.~\ref{fig:figure1}(b)]. 
Note also that, for a given pressure, the percolating cluster coexists with non-percolating clusters of various sizes $s$ and coordination $Z$ [see Fig.~\ref{fig:figure2}(a)].

\begin{figure}[h]
    \centering

    \includegraphics[width=7cm]{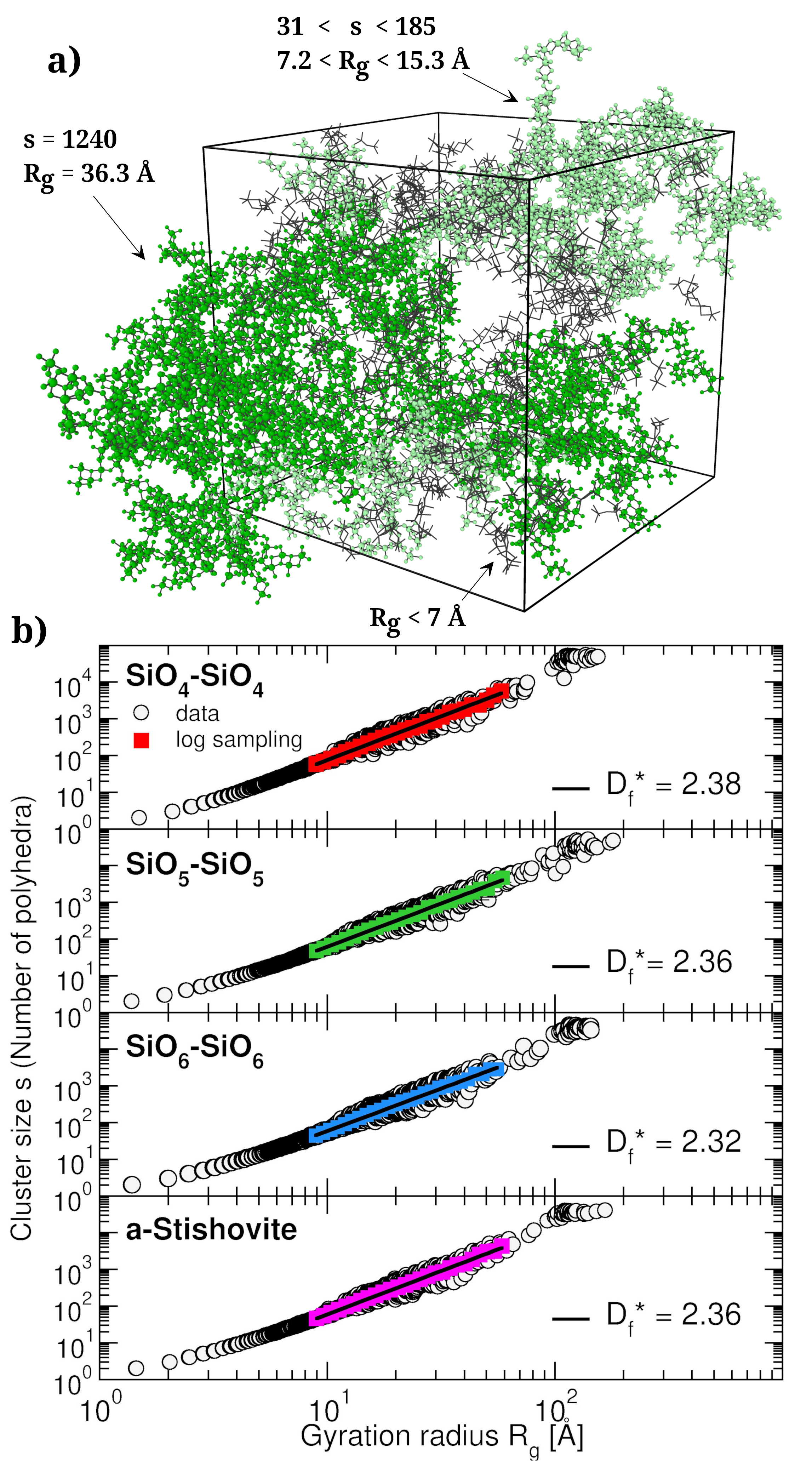}
    \caption{Gyration radius $\textit{R}_\textit{g}$ of SiO$_z$ polyhedra clusters at the critical pressure $\textit{p}_\textit{c}$ for the largest system size. {\bf (a)} Snapshot of unwrapped SiO$_5$-SiO$_5$ clusters at the critical point (27\,k atoms) as an illustration. Light green and dark green colors highlight clusters of increasing gyration radius.    
    Clusters whose $R_g$ is inferior to $7$~\AA\ are shown in gray. {\bf (b)} log-log plot of $s$ versus $R_g$ in the SiO$_z$-SiO$_z$ and stishovite clusters. Open black circles are the raw data, and colored squares are the log-sampled data obtained by dividing the $x$ range into 30 logarithmically spaced bins, then averaging the $y$ values (and propagating errors) within each bin, using bin centers as the new $x$ coordinates, and discarding empty bins. Black lines are fits to the log-sampled data, giving the averaged fractal dimension $D_f^*$. 
    }
    
    \label{fig:figure3}
\end{figure}
\begin{table}[h]
\centering
\begin{tabular}{ccc||ccc|ccc}
\hline 
\hline
\multicolumn{3}{c||}{\textbf{Bonded model}} & \multicolumn{6}{c}{\textbf{Non-bonded model}} \\
\multicolumn{3}{c||}{SiO$_Z$} & \multicolumn{3}{c}{SiO$_2$} & \multicolumn{3}{c}{\hspace{-0.5cm}H$_2$O} \\
\hline
$Z$ & $p_c$(GPa)& $\phi_c$ & $Z$ & $p_c$(GPa) & $\phi_c$ & $Z$ & $p_c$(kBar) &$\phi_c$ \\
\hline
4 & 16.32 & 0.35 & LD & 11.50 & 0.29 & LD & 6.70 & 0.19 \\
5 & 11.86 & 0.29 & HD & 8.25 & 0.28 & HD & 5.60 & 0.28 \\
6 & 23.16 & 0.26 & VHD & 22.50 & 0.21 & VHD & 7.30 & 0.17 \\
6$^*$ & 30.52 & 0.24 & & & & & & \\
\hline
\hline
\end{tabular}
\caption{ Critical pressures $p_c$ and critical fractions $\phi_c$ for the percolation transitions, obtained from the peak of the correlation length $\xi$ for the 1\,056\,000-atom SiO$_2$ and 16\,384-atom H$_2$O glasses, using the SiO$_Z$ descriptor ($a$-SiO$_2$, bonded model), the SiSi$_Z$ descriptor ($a$-SiO$_2$, non-bonded model), and the OO$_Z$ descriptor ($a$-H$_2$O, non-bonded model). 6$^*$ refers to stishovite.}
\label{tab:table_pc}
\end{table}

Figure~\ref{fig:figure2}b shows the pressure dependence of the percolation order parameter $P_\infty$, a quantity that relates to the percolation probability of the polyhedral networks, which is defined by
$$P_{_\infty} = \frac{1}{Q}\sum_{i}\frac{ p_is_i^\text{big}}{N_\text{Si}\phi_i}$$
where $Q$ is the total number of configurations taken into account when averaging and $N_\text{Si}$ is the total number of Si atoms.
For a given $i$ configuration, $p_i$ is zero if no cluster percolates, and $1$ when the spanning cluster becomes infinite along the three dimensions of the simulation box, while $s_i^\text{big}$ is the size of the biggest cluster and $\phi_i$ is the fraction of Si atoms with $Z$ coordination.
We observe similar percolation transitions as those first reported by Hasmy {\it et al.}~\cite{Has2021}, using 
the DFTB approach for a small system size (namely 1008 atoms).
Furthermore, Fig.~\ref{fig:figure2}b reveals that the transition becomes steeper for larger boxes, consistent with the reduced influence of finite-size effects. 
It also shows that the $P_\infty$ data points for (SiO$_6$-SiO$_6$)$_n$ and $a$-stishovite clusters are widely dispersed for boxes containing up to 15\,120 atoms.
The critical pressure $p_c$ of the percolation transitions indicated by vertical dashed lines in Fig.~\ref{fig:figure2}b is defined as the maximum of the cluster correlation function $\xi$ 
(see Fig.~SM1),
which occurs at the onset of the sharp increase in $P_\infty$. 
The step-like shape of the latter indicates the emergence of percolating clusters and delineates the pressure ranges of percolation coexistence: 
(SiO$_4$-SiO$_4$)$_\infty$ and (SiO$_5$-SiO$_5$)$_\infty$ infinite networks between 11 and 16~GPa,  followed by (SiO$_5$-SiO$_5$)$_\infty$ and (SiO$_6$-SiO$_6$)$_\infty$ networks above 21 GPa, to which is added a $a$-stishovite-type structure at pressures above 24~GPa. 
The evolution from low-to-high density polyhedral networks mimics the series of pressurized crystalline counterparts, from deformed quartz-like structures at low pressure,  to coesite IV-, coesite V-, and stishovite-like structures at high pressure \cite{Has2021}.    

Figure~\ref{fig:figure3}a illustrates unwrapped SiO$_5$–SiO$_5$ clusters, with the light green to dark green  ones highlighting clusters corresponding to increasing radii of gyration $R_g$. For the sake of clarity, smaller clusters, in black in the figure, are depicted by their bonds only.
Figure~\ref{fig:figure3}b highlights the power-law relationship  $s \sim R_g^{D_f^*}$ between the cluster size $s$, \textit{i.e.}, the number of polyhedra in the cluster, and the radius of gyration 
$R_g(s)$~\cite{meakin1998fractals}.
The solid lines correspond to the power-law regressions performed on clusters of sizes $R_g \ge 7$~\AA\ ensuring that a fractal dimension is defined appropriately [see illustration in Fig.~\ref{fig:figure3}(a)]. 
The percolating cluster is also  excluded from the analysis.
Accordingly, $D_f^{*}$, which characterizes the dimensionality of a large ensemble of finite clusters, also referred to as ``lattice animals,'' should be regarded as an {\it effective} fractal dimension of the finite-cluster ensemble. It is physically distinct from $D_f$, the fractal dimension of the critical spanning cluster obtained from $S_{\max}\propto L^{D_f}$ at $p_c$. Its value is expected to lie below $D_f$~\cite{Sta2003}.

In addition, as pressure approaches the critical percolation threshold, larger clusters begin to form. 
Finally, as predicted by percolation theory, at the critical percolation threshold, the cluster size distribution follows a power law $n_s \sim s^{-\tau}$, where $\tau$ is called the Fisher exponent 
(see Fig.~SM2).

\subsection{The non-bonded interaction approach}

O'Keeffe and Hyde~\cite{Okeefe1981, Kon2020} developed long ago an alternative approach to describe the structure of non-molecular crystals. 
Their analysis relies on the observation that the distance between {\it non-bonded} first neighbors cations in many molecular crystals is nearly independent of the bridging atom (anion).
Using solely SiSi force constants and assuming rigid regular tetrahedra, they were able to reproduce the bulk modulus of $\alpha$-quartz and $\alpha$-cristobalite \cite{New1980}.
Although distinct from the traditional description, the analysis of the SiSi$_Z$ coordination reproduces the same underlying structural transformations as that of the {\it bonded} approach.
Thus, in crystals both bonded and non-bonded analyses serve as complementary indicators.
However, one approach may be more effective than the other in capturing the structural transformation underlying the change in physical properties.

In ice, each oxygen forms four hydrogen bonds with neighboring oxygens in the first coordination shell in accordance with the Bernal–Fowler ice rules, and this tetrahedral coordination is preserved across a wide range of pressures. As pressure increases, however, some oxygens from the second coordination shell are forced into interstitial positions between the first and second shells, driven by intermolecular non-bonded interactions such as van der Waals forces~\cite{Has2021,Mor2016}.
Consequently, structural transformations are most often assessed using non-bonded criteria  to distinguish low-density (LD) and high-density (HD) amorphous states. Local descriptors such as the coordination number within a cutoff radius can account for this effect, as they can include any interstitial atoms that have migrated between shells.

\begin{figure}[h]
    \centering
    \includegraphics[width=\linewidth]{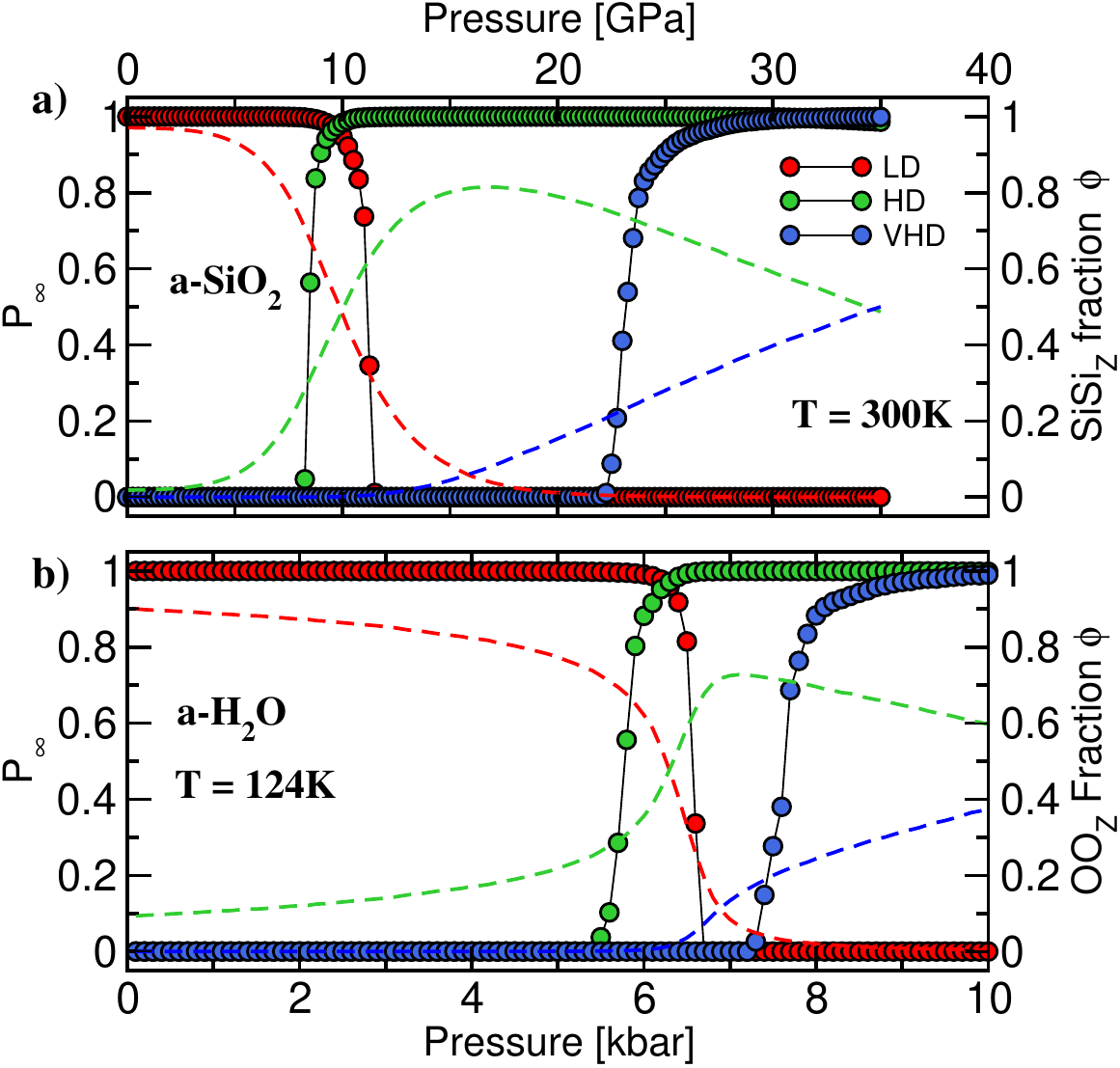}
    \caption{{\bf (a)} $a$-SiO$_2$ and {\bf (b)} $a$-H$_2$O. 
    Circles refer to the order parameter $P_\infty$ and dashed lines to fractions $\phi$. LD stands for $Z=4$, HD for $Z=4$--$7$, and VHD for $Z\geq 8$. 
    \label{fig:figure4}}
\end{figure}
Given our definition of the SiSi$_Z$ local structures in amorphous silica, it remains to be explored whether these properties still apply.
Figure~\ref{fig:figure4}(a) shows the series of percolation transitions of SiSi$_Z$ structures in $a$-SiO$_2$ at $T=300$~K. For comparison, Fig.~\ref{fig:figure4}(b) shows the percolation transitions for OO$_Z$ structures of $a$-H$_2$O at $T=124$~K.  
The pressure dependencies of the fractions of the corresponding  SiSi$_Z$ and OO$_Z$ coordinations are also plotted (dashed lines). The same definitions are adopted in both systems:  LD stands for $Z=4$ (red lines and symbols), HD for $Z=4$--$7$ (red lines and symbols), and VHD for $Z\geq 8$ (blue lines and symbols), as previously proposed for water~\cite{Has2025}.
In \textit{v}-SiO$_2$, the percolation of HD structures occurs at $p_c = 8.25$~GPa, followed by a percolation coexistence of LD and HD states until the LD  network depercolates at $p_c = 11.50$~GPa, {\it i.e.}, the structure no longer spans throughout the box.
Upon further increasing the pressure, a VHD cluster percolates at $p_c=22.50$~GPa. 
The SiSi$_Z$ fractions of the LD and HD ``phases'' intersect in the middle of an LD-HD coexistence region delimited by the percolation of the HD ``phase'' and the depercolation of the LD ``phase.'' 
A similar behavior likely occurs in the HD-VHD domain, but we did not reach high enough pressure to observe the depercolation of the HD structures.   
Interestingly, the HD cluster percolates very close to the transition from the elastic to the plastic regime of silica ($\approx 7$--$8$~GPa in SHIK models), while the VHD cluster emerges in the region of the hysteresis in the elastic properties~\cite{Wei2019} (between $\approx 20$--$25$~GPa).   These aspects  will be addressed in detail in a forthcoming paper.

Within the non-bonded approach, Fig.~\ref{fig:figure4} also shows how the structural evolutions under pressure of samples of SiO$_2$ glass at room temperature and amorphous ice, at 124~K, follow the same sequence of percolation transitions, but occurring at pressures about 20 times lower in ice.
In the latter, the LD-HD transition is associated with a peak in the compressibility and a sharp increase in the density, which underscores the close connection between percolation and physical properties~\cite{Has2025}.  

\begin{figure}[h]
    \centering
    \includegraphics[width=\linewidth]{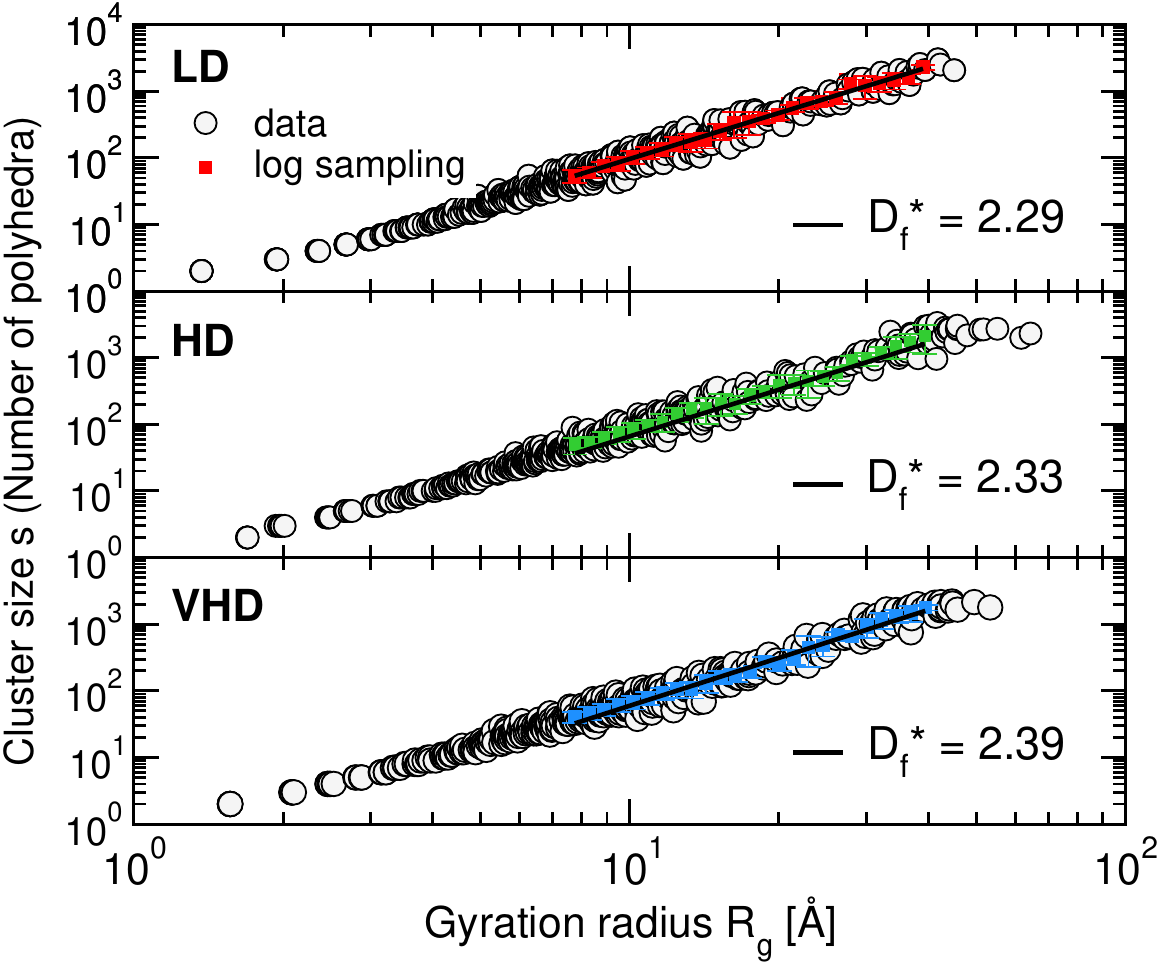}
    \caption{Gyration radius $R_g$ of LD, HD, and VHD  clusters at the critical pressure $p_c$. Lines are the fits with a power law, $s\propto R_g^{ D_f^*}$.
   } 
    \label{fig:figure5}
\end{figure}

The fractal dimensions $D_f^*$ of the LD, HD, and VHD ``phases'' in $a$-SiO$_2$ and $a$-H$_2$O extracted from the distribution of the gyration radius, $s\propto R_g^{D_f^*}$ (see Fig.~\ref{fig:figure5}), have  been  calculated and the values are summarized in Table~\ref{tab:critical_exponents}. 

\begin{figure}[b]
    \centering
    \includegraphics[width=\linewidth]{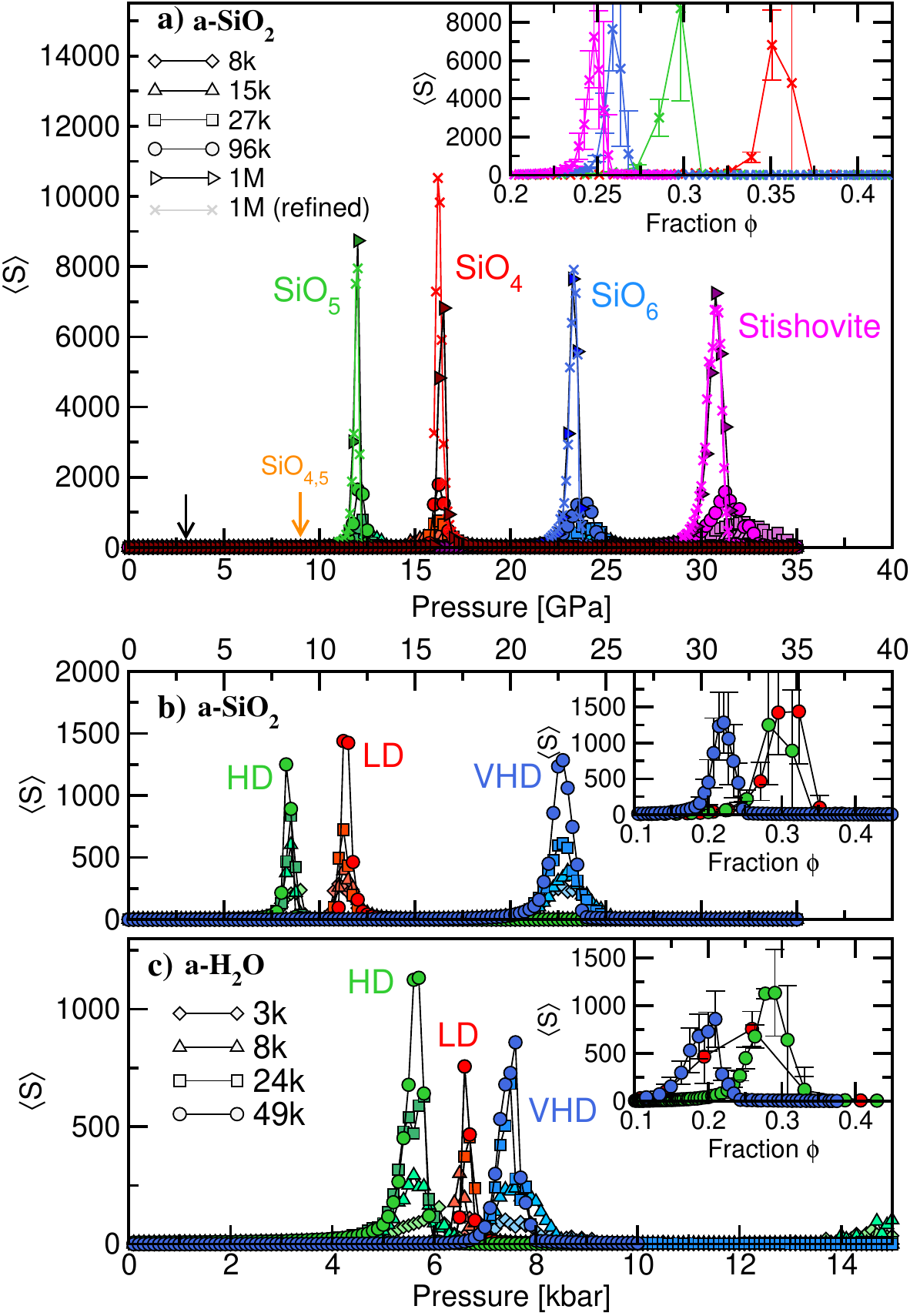}
    \caption{{Average cluster size $\langle S\rangle$ as a function of pressure and fraction (inset)}. {\bf (a)} bonded approach in $a$-SiO$_2$ (SiO$_Z$-SiO$_Z$ and stishovite clusters). The black arrow indicates the minimum of the bulk modulus, while the orange one marks the alternated SiO$_4$-SiO$_5$ clusters percolation transition~\cite{perradin_polyamorphism_2025}. {\bf (b)} non-bonded approach in $a$-SiO$_2$ (SiSi$_Z$-SiSi$_Z$ LD, HD, and VHD clusters) and {\bf (c)} non-bonded approach in $a$-H$_2$O for LD, HD, and VHD clusters.
    \label{fig:figure6}}
\end{figure}

\section{Finite-size scaling and critical exponents}

The average distance between two sites within the same cluster, known as the connectivity correlation length $\xi$, quantifies the range over which fluctuations in the system are correlated near the percolation critical point. 
This quantity follows a scaling behavior described by the critical exponent as $\xi(\phi) \sim |\phi - \phi_c|^{-\nu}$, where $\phi$ is the occupation probability and $\phi_c$ is its critical percolation value when $P_\infty$ becomes 1~\cite{meakin1998fractals}.
Here, $\phi$ is the fraction of polyhedra with coordination $Z$, which correlates with the system's pressure $P$.
Indeed, as the system approaches the percolation threshold, $\xi$ diverges toward infinity in the thermodynamic limit 
(see Fig.~SM1).

This divergence shows that, at the percolation critical point, fluctuations extend over all length scales, giving rise to the scaling behavior of the order parameter $P_{\infty} \sim (\phi - \phi_c)^{\beta}$, or similarly $P_{\infty} \sim (P - p_c)^{\beta}$, reflecting the universal features of a second-order phase transition. 
Similarly, the average cluster size $\langle S \rangle$ follows a scaling law,  $\langle S \rangle \sim |\phi - \phi_c|^{-\gamma}$ in the vicinity of the percolation critical point~\cite{Sta2003}.
The critical exponents $\nu$, $\beta$ and $\gamma$ are related to those of thermal phase transitions, and the hyperscaling relation
$\nu d = 2\beta + \gamma$
holds for percolation as well. 
By analogy with magnetic systems, the second moment $\langle S\rangle$ is related to the magnetic susceptibility $\chi$, while the infinite cluster probability $P_{\infty}$ corresponds to the magnetic order parameter $M$~\cite{Sta2003}.

Figure~\ref{fig:figure6} shows $\langle S\rangle$ as a function of pressure $P$ in the main panel and as a function of the SiO$_Z$ fraction $\phi$ in the inset, for the different system sizes. 
The divergence is much sharper in larger boxes, highlighting box-size effects and the difficulty of obtaining accurate values for the critical exponents directly from the power laws.  
In three-dimensional regular lattices, the critical fraction $\phi_c$ of percolating units, where $\xi$ (or $\langle S \rangle$) exhibits a maximum, decreases with increasing coordination number $Z$ of the network~\cite{Gau1983}.
A similar trend is observed in $a$-SiO$_2$ as shown in the inset of Fig.~\ref{fig:figure6}a. For regular networks, computational studies predict $\phi_c(p_c) = 0.4299$ for the tetrahedral arrangement (diamond structure) and $\phi_c(p_c) = 0.3117$ for the octahedral configuration (simple cubic structure) for site percolation~\cite{Cre1992}. 
Despite the presence of structural disorder, our calculations in $a$-SiO$_2$ yield values that are quite close: $\phi^{\text{SiO}_4}_c(p_c) \simeq 0.39$ and $\phi^{\text{SiO}_6}_c(p_c) \simeq 0.27$, respectively. 

In Fig.~\ref{fig:figure6}b, we compute the average cluster size of SiSi$_Z$ clusters defined in the non-bonded approach. 
The divergence of the LD, HD, and VHD peaks is also quite sharp here, mirroring the results obtained with SiO$_z$-based structural analysis.
This behavior contrasts with that observed for the non-bonded approximation of $a$-H$_2$O [shown in Fig.~\ref{fig:figure6}(c)], where the peaks are broader both for the pressure and fraction $\phi$ dependences. 
The estimated values of $\phi_c(p_c)$ for different coordination numbers $Z$ in both \textit{a}-SiO$_2$ and \textit{a}-H$_2$O are given in Table~\ref{tab:table_pc}. 
We notice that these values are in close agreement with computational studies and percolation models on crystalline lattices~\cite{Sta1979,Cre1992}.

\begin{figure*}
    \centering
    \includegraphics[width=0.8\linewidth]{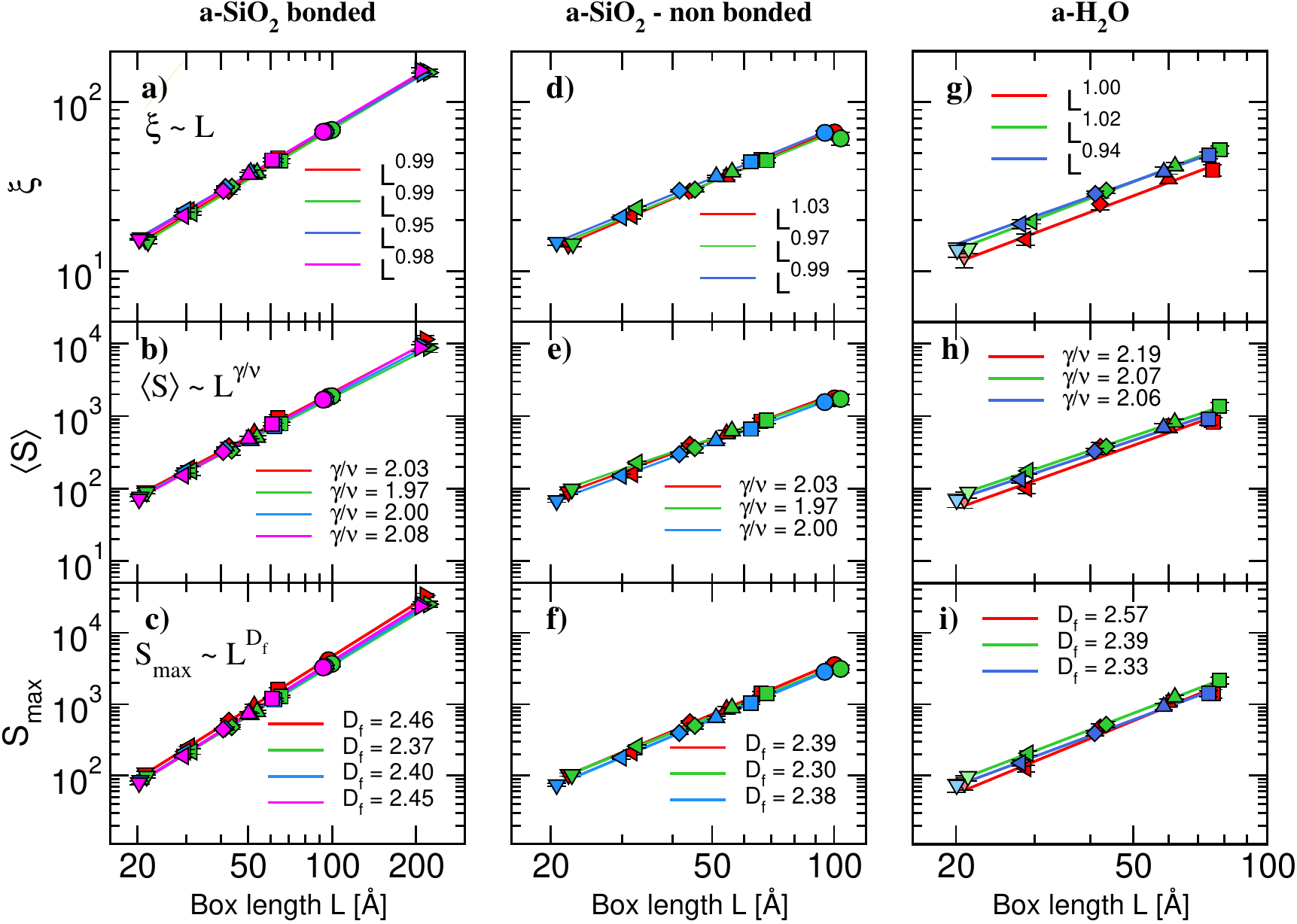}

    \caption{Scaling laws of the correlation length $\xi$, the average cluster size $\langle S\rangle$, and the spanning cluster size 
    $S_{max}$ at $p_c$.  \textbf{(a)}, \textbf{(b)}, and \textbf{(c)} for $a$-SiO$_2$ bonded,  \textbf{(d)}, \textbf{(e)}, and \textbf{(f)} for $a$-SiO$_2$ non-bonded, and \textbf{(g)}, \textbf{(h)}, and \textbf{(i)} for $a$-H$_2$O. 
    Symbols refer to box size and colors to $a$-SiO$_Z$ clusters, $Z$=4 (red), $Z$=5 (green), $Z$=6 (blue), $Z$=6$^*$ (magenta), and $a$-SiO$_2$ non-bonded, $a$-H$_2$O clusters LD (red), HD (green), and VHD (blue). Lines are power-law fits and the obtained exponents
    are given in the legends.}
    \label{fig:figure7}
\end{figure*}
Prior to applying the finite-size scaling ansatz to calculate the critical exponents, it is important to note that, in finite-size systems, the correlation length $\xi$ cannot exceed the simulation box size $L$.  Consequently, near criticality $\xi$ should scale with $L$, i.e., $\xi\propto L^a$ with $a=1$. 
This condition is fairly well satisfied for all polyhedra clusters of $a$-SiO$_2$ and $a$-H$_2$O, as shown in Fig.~\ref{fig:figure7}a,d and g.

The validity of this scaling law is a prerequisite for the analysis below.
When this condition is satisfied, the correlation length at $p_c$ reads $\xi (p_c,L)\propto L$ and replacing $\xi$ by $L$ in the divergence  $\xi \propto |P-p_c|^{-\nu}$ yields 
$$|P-p_c|\propto \xi^{-1/\nu}\propto L^{-1/\nu}$$
As a consequence, the scaling laws followed by the  divergences of the other percolation properties are
\begin{align}
    P_\infty(p_c,L) \propto L^{-\frac{\beta}{\nu}} \label{eq:fss_order_paramter} , \\
    \langle S\rangle(p_c,L) \propto L^{\frac{\gamma}{\nu}} \label{eq:fss_average_size} ,\\
    S_{\text{max}}(p_c,L) \propto L^{D_f} ,\label{eq:fss_spanning_size}
\end{align}
where $D_f$ is the fractal dimension of the largest cluster precisely at $p_c$.
For each percolation transition, $p_c$ is not assigned a single fixed value but is determined for each sample from the peak position of $\xi$, $\langle S\rangle$, or $S_{\max}$, respectively. The peak values and the corresponding box lengths are then averaged over the five independent samples for each system size, yielding one $p_c(L)$ value for each observable and each $L$ (hereafter referred to as the “peak-based” method).
The scaling behavior of the average cluster size $\langle S\rangle (p_c)$ and of the largest cluster $S_{max} (p_c)$ for amorphous silica (bonded approach) and ice, is shown in Fig.~\ref{fig:figure7}. The fits yield $\gamma/\nu$ and $D_f$, respectively. 
Estimating $\beta/\nu$ from $P_\infty$ scaling is usually very demanding because it requires extensive sampling and high statistical precision. 
This can be observed directly in Fig.~\ref{fig:figure2}b where the variation of $P_\infty$ at $p_c$ for the different box size $L$ is rather weak compared with its spread. In contrast, the variation of the other quantities, like $\langle S\rangle$ in Fig.~\ref{fig:figure6} and $\xi$ in Fig.~SM1, are large close to $p_c$. 
Given the above, we could not obtain reliable values of the  $\beta/\nu$ ratio from the scaling of the order parameter $P_\infty$.

\section{Discussion}

\subsection*{Bonded vs non-bonded model}

\begin{table*}
\centering
    \begin{ruledtabular}

    \renewcommand{\arraystretch}{1} 
    \begin{tabular}{ccccccccccccc}
    Power or scaling & \textbf{Perco} & \textbf{Ising} 
    & \multicolumn{4}{c}{\textbf{SiO$\mathbf{_Z}$-SiO$\mathbf{_Z}$}} &
    \multicolumn{3}{c}{\textbf{SiSi$\mathbf{_Z}$-SiSi$\mathbf{_Z}$}} & \multicolumn{3}{c}{\textbf{H$_2$O}} \\
    laws & \textbf{(3D)} & \textbf{(3D)} 
    & $\boldsymbol{Z=4}$ & $\mathbf{Z=5}$ & $\mathbf{Z=6}$ & $\mathbf{Z=6^*}$  & \it{\textbf{LD}} & \textbf{HD} & \textbf{VHD} & \it{\textbf{LD}} & \textbf{HD} & \textbf{VHD} \\
    \hline
 $M \propto R_g^{D_f^*}$ & \textit{-} & - & \it{2.38}(5) & 2.36(6) & 2.32(5) & 2.36(6) & \it{2.31}(7) & 2.33(7) & 2.39(6) & \it{2.29}(7) & 2.33(10) & 2.39(7) \\

    $n_s\propto s^{-\tau}$  & 2.18 & - & \it{2.12}(1) & 1.92(1) & 1.90(1) & 1.93(1) & \it{2.14}(2) & 1.98(3) & 1.90(2) & \it{2.38}(4) & 2.00(7) & 1.83(4) \\
   \hline
    $\xi \propto L^a$  & 1.0 & 1.0 & \it{0.99}(1) & 0.99(2) & 0.95(2) & 0.98(1) & \it{1.03}(3) & 0.97(4) & 0.99(3) & \it{1.00}(10) & 1.02(5) & 0.94(7) \\

     $\langle S\rangle \propto L^{\gamma/\nu}$ & 2.05 & 1.97 & \it{2.03}(5) & 1.97(5) & 2.00(8) & 2.08(4) & \it{2.06}(7) & 1.92(7) & 2.05(4) & \it{2.19}(14) & 2.07(11) & 2.06(13) \\
     
    $S_\text{max}\propto L^{D_f}$ & 2.53 & - & \it{2.46}(4) & 2.37(5) & 2.40(5) & 2.45(3) & \it{2.39}(6) & 2.30(6) & 2.39(4) & \it{2.57}(14) & 2.39(11) & 2.33(13) \\ 

   \end{tabular}
    \caption{
     Critical exponents obtained for $a$-SiO$_2$ within the bonded approach (SiO$_z$-SiO$_z$ structures) and the non-bonded approach (SiSi$_z$-SiSi$_z$ structures and stishovite, $Z=6^*$), and in $a$-H$_2$O  within the non-bonded approach. The entries in italics correspond to depercolation processes.
     The values for standard (3D) percolation and for the Ising (3D) model are also shown for comparison~\cite{Sta1979}. 
    The first two rows report power-law fits obtained for the 10$^6$-atom silica sample (respectively, the 49\,k-atom ice sample). 
     The last three rows are exponents resulting from the scaling law derived from the finite-size scaling ansatz. 
     Values are given as the mean with uncertainty in parentheses indicating the error on the last digit(s).
     }
    \label{tab:critical_exponents}
    \end{ruledtabular}
\end{table*}

Our results indicate that bonded and non-bonded approaches provide complementary insights, offering a particularly new understanding of the intermediate-pressure regime in compressed SiO$_2$ glass, a region for which no crystalline counterpart has yet been identified, i.e., up to the onset of (SiO$_6$–SiO$_6$)$_\infty$ corner-sharing percolation.

Moreover, comparing these two approaches provides a unifying framework for elucidating the structural transformations and common origins of the pressure-induced anomalies shared by bonded glasses, such as SiO$_2$, and non-bonded glasses, such as amorphous ice. 
For SiO$_2$ at low pressures, both approaches exhibit a tetrahedra-dominated network until the bulk modulus reaches a minimum around 3~GPa [black arrow in Fig.~\ref{fig:figure6}(a)], suggesting local distortion of the tetrahedral units associated with the emergence of SiO$_5$ pentahedra~\cite{perradin_polyamorphism_2025}. 
For pressures above the minimum,  
Fig.~\ref{fig:figure6} reveals a coherent picture between the two approaches: the high-density (HD) cluster begins to percolate concurrently with the appearance of alternating (SiO$_4$-SiO$_5$)$_\infty$ connections in the bonded description [orange arrow in Fig.~\ref{fig:figure6}(a)], just above the bulk modulus minimum~\cite{Has2021}, while the low-density (LD) cluster depercolates at the pressure where (SiO$_5$-SiO$_5$)$_\infty$ percolation sets in (see Fig.~\ref{fig:figure6}). Over this same pressure range, (SiO$_4$-SiO$_4$)$_\infty$ and (SiO$_5$-SiO$_5$)$_\infty$ percolating clusters coexist, as reported previously~\cite{Has2021}. This picture is consistent with earlier suggestions that percolating clusters containing mixtures of SiO$_4$ and SiO$_5$ units in this intermediate regime are reminiscent of the pressure-induced post-quartz amorphous states~\cite{Hu2017,Has2021}.

Beyond 13~GPa, an infinite (SiO$_6$-SiO$_6$)$_\infty$ cluster emerges and percolates at the same pressure where VHD percolating cluster emerges, as shown in Fig.~\ref{fig:figure6}. Within the bonded approach, the coexistence of all SiO$_Z$ percolating clusters beyond 13~GPa bears a close structural resemblance to coesite-IV, whose crystalline structure combines  SiO$_4$ and SiO$_5$ units alongside emerging (SiO$_6$-SiO$_6$)$_\infty$  percolating structures. 
Beyond 17~GPa, SiO$_4$ cluster depercolates, and the amorphous structure increasingly resembles coesite-V, in which SiO$_5$ and SiO$_6$
units dominate. This structural analogy persists up to $\sim$30~GPa, where the connectivity pattern becomes reminiscent of the stishovite polymorph.
An additional advantage of the non-bonded picture is that it enables direct comparison with non-bonded glasses, such as amorphous H$_2$O. As shown in Fig.~\ref{fig:figure6}c, glassy water exhibits the same sequence of percolation transitions, albeit shifted to different pressures owing to substance-specific interactions.

\subsection*{Percolation criticality}

Critical exponents obtained from the scaling laws and their uncertainties were estimated using two independent procedures, the standard error-propagation approach and a bootstrap analysis that also accounts for uncertainties on the critical pressure $p_c$ (see Supplemental Material~\cite{supmat}). Both methods yield similar central values and uncertainties for the critical exponents. 
The latter are summarized in Table~\ref{tab:critical_exponents} with the error on the last digit given in parentheses.
For reference, the expected values for standard 3D percolation and for the Ising model are also included.
For all structural descriptors considered in this study, the correlation length scales with the box size as $\xi\propto L^a$ with $a\simeq 1$, which is consistent with critical behavior. 
The genuine critical regime of transitions near the pressure threshold is further confirmed by the cluster size distribution $n_s$ that scales with $n_s\propto s^{-\tau}$ over almost two orders of magnitude (see Fig.~\,SM2). 
Therefore, the nature of the universality class merits discussion, particularly in light of the critical exponents. 
We emphasize that the compressed configurations are out-of-equilibrium, history-dependent glassy states, and that the increase in the relaxation times of the glass with pressure~\cite{Ono2020} progressively reduces the effective statistical sampling of the percolation observables. The percolation criticality reported here is therefore geometrical rather than thermodynamic and does not require full ergodic equilibration. We consequently regard its possible association with a distinct universality class as a physically motivated hypothesis rather than a definitive identification.

Let us consider first the case of amorphous silica. 
It is striking that the Fisher exponent $\tau$ for depercolation of the SiO$_4$-SiO$_4$  and SiSi$_4$-SiSi$_4$ (LD) tetrahedra clusters  (2.12 and 2.14, respectively) are much  closer to the value expected for standard percolation (2.18) than for the percolation of other polyhedral networks, in either the bonded or non-bonded approach.
It should be noted, however, that the exponents reported here are effective, window-dependent estimates
rather than asymptotic critical exponents. They are obtained from a power-law fit of
$n_s\propto s^{-\tau}$ (Fig.~SM2) that extends to small cluster sizes ($s\gtrsim 2$),
where non-universal corrections flatten the distribution and lower the apparent $\tau$,
while at large $s$ the scaling cutoff and the limited statistics act in the opposite
direction. 
The estimate is therefore sensitive to the fitting window and to the location of
$p_c$. In particular, values $\tau<2$ cannot be asymptotic percolation exponents, since
$\tau = 1 + d/D_f \ge 2$ for $D_f\le d$.

In all cases, the fractal dimension $D_f$ is systematically lower than the reference value of 2.53. 
Because $D_f$ is tied to the largest cluster at $p_c$, it is particularly sensitive to the precise location of $p_c$ and an imprecise estimate of the latter can significantly bias $D_f$. 
To assess whether this discrepancy reflects genuine universality classes, we determined, for each percolation transition, the asymptotic critical pressure in the thermodynamic limit, $p_c^\infty$. We then compared the critical exponents obtained from scaling analyses performed at fixed $p_c^\infty$ with those obtained using the peak-based procedure employed in Table~\ref{tab:critical_exponents}. 
The results summarized in the Supplemental Material~\cite{supmat} show that a better estimate of $p_c$ substantially improves the agreement of $D_f$ with the percolation value  across all four transitions (2.58, 2.48, 2.44, and 2.51; see Table~SM5), as well as the consistency with the hyperscaling relation (Table SM6). 
Moreover, the discrepancy observed when using the peak-based estimate of $p_c$ goes in the same direction as the hyperscaling mismatch, {\it i.e.}, it is consistent with a one-sided bias in the estimate of $p_c$ rather than with a distinct universality class.

Finally, the ratios $\gamma/\nu$ are all very close to the reference value of 2.05.
Unfortunately, this does not tell whether the individual values match those of standard percolation, $\gamma=1.80$ and $\nu=0.88$.

The finite-size data collapse of the mean cluster size (Fig.~SM4) offers a
complementary view by estimating $\nu$ and $\gamma$ separately. 
The fitting procedure reveals appreciable correlations between the two exponents, indicating that a more robust analysis would require more extensive statistical sampling, particularly over a range of box sizes spanning more than one order of magnitude. Nevertheless, a general trend can still be identified.
For the lower-pressure transitions, i.e., the
$\mathrm{(SiO_4})_\infty$ network depercolation and the $(\mathrm{SiO_5})_\infty$ network percolation, the free collapses yield $\nu \approx 0.85$ and
$\gamma \approx 1.8$, consistent with ordinary three-dimensional percolation
($\nu \approx 0.88$, $\gamma \approx 1.80$) and incompatible with other classes
such as, for example, the Ising or Fortuin--Kasteleyn class~\cite{FortuinKasteleyn1972}
($\nu \approx 0.63$), for which the collapse fails markedly. 
This matches the expectation that short-range structural correlations do not modify the
ordinary-percolation exponents.
At higher pressure
[$(\mathrm{SiO_6})_\infty$ and stishovite-like networks], the free-collapse
exponents drift away from the percolation values (i.e., $\nu \approx 1.13$ for the
$\mathrm{SiO_6}$ network). We stress, however, that the $\gamma/\nu$ ratios implied
by these free collapses do not fully coincide with those obtained directly from the
$\langle S\rangle \propto L^{\gamma/\nu}$ finite-size scaling using the peak-based method (Table~\ref{tab:critical_exponents}), but are consistent, within errors, with those derived using $p_c$ fixed to $p_c^\infty$ (Table~SM6).
The separate determination of $\nu$ and $\gamma$
is thus not tightly constrained, and we use the collapses only qualitatively.
With this caveat, the high-pressure drift admits two non-exclusive interpretations. It may reflect the emergence of long-range spatial correlations in the increasingly coordinated and mechanically constrained network, as elastic properties are intrinsically long-ranged. Such longer-ranged correlations could modify the critical exponents, with $\nu > 0.88$. Alternatively, the drift may result from a reduction in the effective statistical quality of the data at high pressure. \\
Combining the exponents in Table~\ref{tab:critical_exponents}, the data collapse ($\nu\approx 0.85$) and
the analysis at fixed $p_c^\infty$ ($D_f$ and $\gamma/\nu$, with the closure of the
hyperscaling relation), there is strong evidence that the low-pressure transitions, namely, the
SiO$_4$ depercolation and the SiO$_5$ percolation, belong to the ordinary
three-dimensional percolation universality class and are incompatible with classes such
as Ising. 
At higher pressure, the collapse suggests a departure from ordinary percolation, even
further from the Ising class, although $D_f$ evaluated at $p_c^\infty$ and $\gamma/\nu$ 
remain compatible with it. The characterization of this possible departure requires
further refinement.\\
The larger box sizes of the $a$-H$_2$O models considered in this study are much smaller than those of $a$-SiO$_2$ and, accordingly, the errors on the exponents are significantly larger. Nevertheless, Table~\ref{tab:critical_exponents} reveals a trend similar to that observed for $a$-SiO$_2$.
The exponents associated with the LD depercolation transition are closer to those of standard percolation than the exponents associated with the HD and VHD clusters.   

\section{Conclusion}

Taken together, these results underline the consistency between the bonded and non-bonded approaches.
The low- and moderate-pressure transitions, including the depercolation of the native tetrahedral SiO$_4$ and LD clusters and the percolation of  SiO$_5$ clusters, occur within a weakly heterogeneous environment and proceed through a process close to random percolation, as suggested by the corresponding critical exponents.
The possibility that this behavior is a general feature associated with a weakly correlated or uncorrelated cluster-growth process in glasses remains to be explored.
For high-pressure percolation transitions, i.e., involving clusters with high coordination number and connectivity, 
the data collapse suggests a possible departure from ordinary percolation, although the values of $D_f$ evaluated at $p_c^\infty$ and $\gamma/\nu$ remain compatible with the corresponding standard values. Establishing such a departure would require more extensive sampling.
It can be argued that for these structures, the transition occurs in a medium where the percolating cluster is surrounded by an infinite cluster of lower coordination and connectivity, alongside emerging clusters with higher coordination and connectivity, resulting in topological, and hence elastic, heterogeneities.
This proposed behavior qualitatively recalls topological constraint theory, also known as percolation rigidity, that arises from a flexible to a rigid network as local connectivity changes~\cite{Tho1985,Tho2002}. 
Furthermore, it has been proposed that the glassy states are associated with an energy megabasin that includes a deep minimum of a crystalline polymorph~\cite{McM2021}.
In $a$-SiO$_2$ the megabasins could correspond to the coesite IV, coesite V, and stishovite phases proposed in Ref.~\cite{Has2021} and in this study.
Around a deep minimum, the polymorph loses long-range order but retains its local structure or connectivity, with a broad local density distribution due to disorder.

From this perspective, tracking percolation-driven structural changes under varying thermodynamic conditions ($P$, $T$) allows for the evaluation of transitions between distinct megabasins, thereby providing a basis for proposed analogies between amorphous structures and crystalline polymorphs~\cite{Has2021,Has2025, Tan2024}. Within this framework, one can also address longstanding issues, such as the origin of plasticity and the mechanical properties of glasses.

\section*{Acknowledgments}

This work was granted access to the HPC resources of IDRIS and CINES under allocations 2023-[A0140910788], 2024-[A0160910788], and 2025-[AD010910788R1], made by GENCI.
J.P. acknowledges financial support from the French Ministry of Higher Education and Research (MESR) through a doctoral fellowship.

\section*{Data availability}

There are no publicly available research data or software supporting this paper. Requests for further information or data should be sent to the authors.

\nocite{binder_monte_2010,Newman_Barkema_1999}
\bibliography{main}

\renewcommand{\theequation}{S\arabic{equation}}
\renewcommand{\thepage}{S\arabic{page}}

\setcounter{equation}{0}
\setcounter{page}{1}
\setcounter{figure}{0}                 
\renewcommand{\thefigure}{SM\arabic{figure}}
\renewcommand{\thetable}{SM\arabic{table}}
\setcounter{table}{0}                 
\renewcommand{\thetable}{SM\arabic{table}}
\clearpage
\onecolumngrid
\raggedbottom       
\begin{center}
{\Large \textbf{Supplemental Material for:}}\\[0.3em]
{\large \textbf{Percolation criticality of amorphous-amorphous transitions in compressed glasses}}\\[1em]
J. Perradin$^{1}$, S. Ispas$^{1}$, R. Paredes$^{2,3}$, A. Hasmy$^{1,4}$, and B. Hehlen$^{1}$\\[0.5em]
{\small
$^{1}$ Laboratoire Charles Coulomb (L2C), CNRS–Université de Montpellier, 34095 Montpellier, France\\
$^{2}$ Departamento de Física y Matemáticas, Universidad Iberoamericana, 01219 Ciudad de México, Mexico\\
$^{3}$ Centro de Física, IVIC, Apdo. 21827, 1020A Caracas, Venezuela\\
$^{4}$ Departamento de Física, Universidad Simón Bolívar, Valle de Sartenejas, Caracas, Venezuela
}
\end{center}

\medskip
\, \\

\noindent {\bf Cluster analysis}\\
To characterize the percolation transition, the percolation probability $\Pi$ and the order parameter $P_{_\infty}$ of the SiO$_z$-SiO$_z$, stishovite, LD, HD, and VHD clusters are estimated using the expressions: 
\begin{equation}\Pi=\frac{\sum_i p_i}{Q},\end{equation}
and,

\begin{equation}P_{_\infty} = \frac{1}{Q}\sum_{i}\frac{ p_is_i^\text{big}}{N\phi_i}.\end{equation}
The sums run over the total number $Q$ of averaged configurations. 
For a given $i$ configuration, $p_i$ is 0 if any cluster percolates, and 1 otherwise. $N$ denotes the total number of SiO$_2$ and H$_2$O molecules and $s_i^\text{max}$ the size of the largest cluster.
$\Pi$ and $P_{_\infty}$ were estimated for three different situations, when the largest cluster percolates in three dimensions. 
The average cluster size $S$ and the correlation length $\xi$ were calculated using~\cite{Sta2003}:
\begin{equation}\langle S\rangle={\frac{\sum_ss^2n_s}{\sum_s sn_s}}\end{equation}
and,
\begin{equation}\xi^2=\frac{\sum_s 2R_{g,s}^2s^2n_s}{\sum_s s^2n_s},\end{equation}
where $R_{g,s}$  is the gyration radius and $n_s$ is the number of clusters of size $s$.
The sums run over all clusters of size $s$, excluding the largest one if it percolates.
The gyration radius $R_{g,s}$ was estimated using the relation:
\begin{equation}R_{g,s}^2= \frac{1}{2s^2}\sum_{i,j} r_{ij}^2,\end{equation}
where the sum runs over pairs of O atoms belonging to each cluster of size $s$. 

$P_{\infty}$ corresponds to a measure of the percolation transition, and near the critical point it exhibits a behavior that obeys a scaling law with a critical exponent for $p<p_c$, as expected for an order parameter~\cite{Sta2003}: 

\begin{equation}
    P_\infty (p) \propto (P-p_c)^\beta,
    \label{eq:scaling_law_pinf}
\end{equation}
and $P_\infty =0$ for $P<p_c$.  

The correlation length $\xi$ measures the distance over which fluctuations in a system are correlated. As a system approaches the critical point, $\xi$ tends to infinity (in the thermodynamic limit). This divergence implies that at the critical point, fluctuations occur at all length scales. Therefore, in the vicinity of the critical point, $\xi$ follows the power law behavior:

\begin{equation}\xi(p) \sim |P - p_c|^{-\nu}.\end{equation}

Similarly, the average cluster size $\langle S\rangle$ also displays a power-law type behavior when approaching $p_c$: 

\begin{equation}\langle S\rangle(p) \sim |P - p_c|^{-\gamma}.\end{equation}

\begin{figure}
    \centering
    \includegraphics[width=8cm]{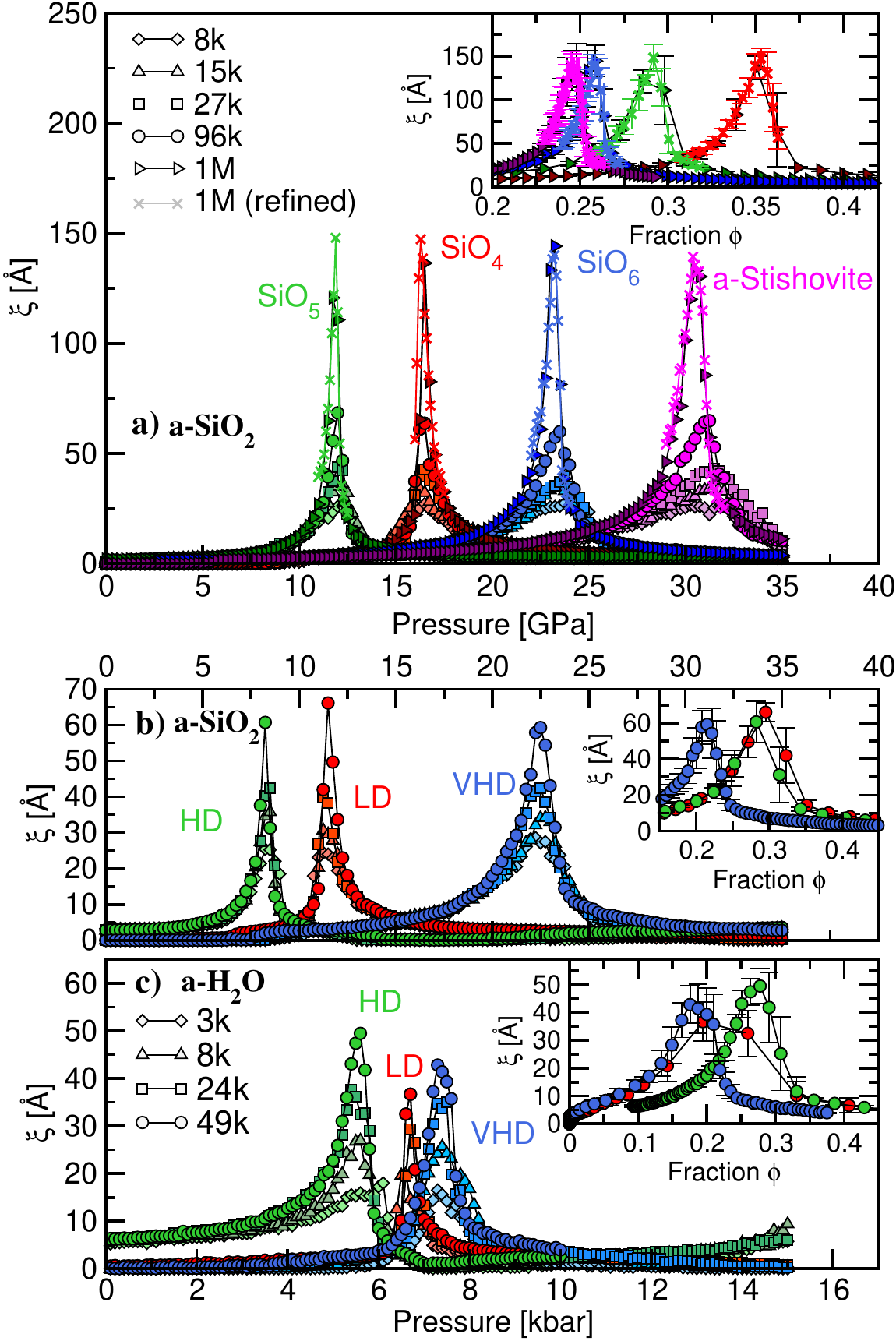}
    \caption{ {\bf Correlation length $\xi$ as a function of pressure and fraction (inset)}. {\bf a)} bonded approach in $a$-SiO$_2$ (SiO$_z$-SiO$_z$ and stishovite clusters) {\bf b)} non-bonded approach in $a$-SiO$_2$ (SiSi$_z$-SiSi$_z$ LD, HD, and VHD clusters) and {\bf c)} non-bonded approach in $a$-H$_2$O for LD, HD, and VHD clusters.}
    
    \label{fig:SM1}
\end{figure}


\begin{figure}
\centering
    \includegraphics[width=8cm]{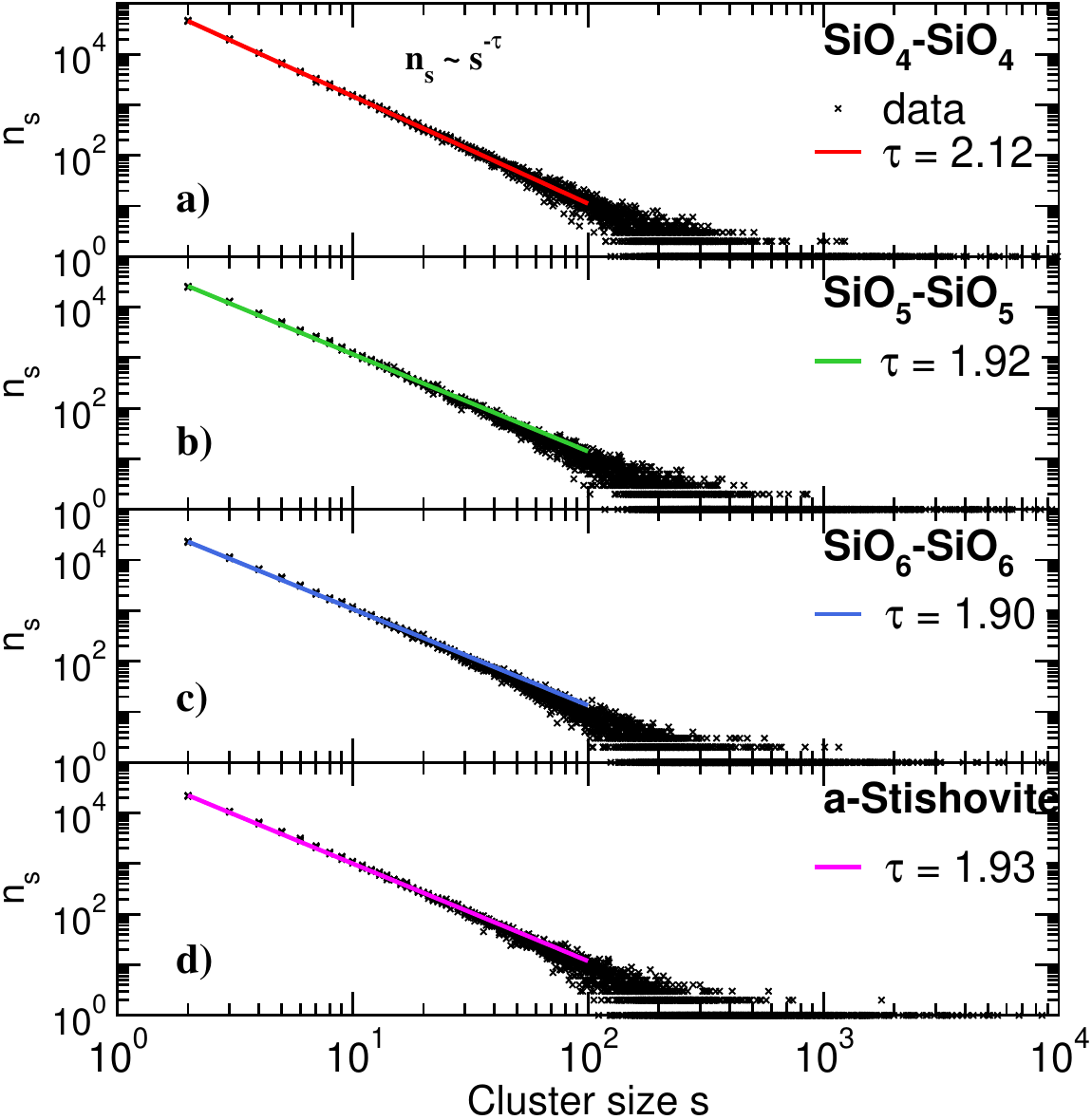}
    \caption{ Cluster size distribution of \textit{a}-SiO$_2$ at the critical pressure $p_c$ for the system size with one million atoms, and fit of the Fisher exponent $\tau$ in the region $s\in[2,100]$ using a power law: \textbf{a)} SiO$_4$-SiO$_4$, \textbf{b)} SiO$_5$-SiO$_5$, \textbf{c)} SiO$_6$-SiO$_6$, and \textbf{d)} stishovite-like clusters.} 
    
    \label{fig:SM2}
\end{figure}

{\bf Data collapse}\\

Applying scaling relations with respect to the simulation box size $L$ at the critical thresholds yields ratios of the different exponents, rather than the individual values of $\nu$, $\gamma$, and $\beta$.
The latter can be obtained by coming back to the reference form of the FSS ansatz. When $\xi$ becomes comparable to $L$, the ansatz proposes that any singular observable $A_L$ (e.g., $\langle S\rangle$ or $P_\infty$) measured in a finite system size in the critical region near $p_c$, collapses in a universal scaling form~\cite{binder_monte_2010}:
\begin{equation}
A_L(\rho) = L^{\zeta/\nu}~f(L^{1/\nu}(\rho-\rho_c)) ,
    \label{eq:fssa}
\end{equation}
where $\zeta$ is the critical exponent related to $A_L$ such as $\gamma$ or $\beta$, respectively, and $\rho$ the pressure $p$ or the SiO$_z$ fraction $\phi$.

Figure~\ref{fig:SM4} shows the result for the four types of polyhedra clusters defined within the bonded approach, and highlights the limitation of the process. 
The collapse obtained using the theoretical exponents (right panels) is very similar to that resulting from a fit (middle panels), even though exponent values can be significantly different (see e.g. SiO$_6$ and $a$-Stishovite clusters on the third and fourth rows, respectively). This partly arises from the data's large error bars, which lead to correlated sets of exponents.

\begin{figure}
    \centering
    \includegraphics[width=8cm]{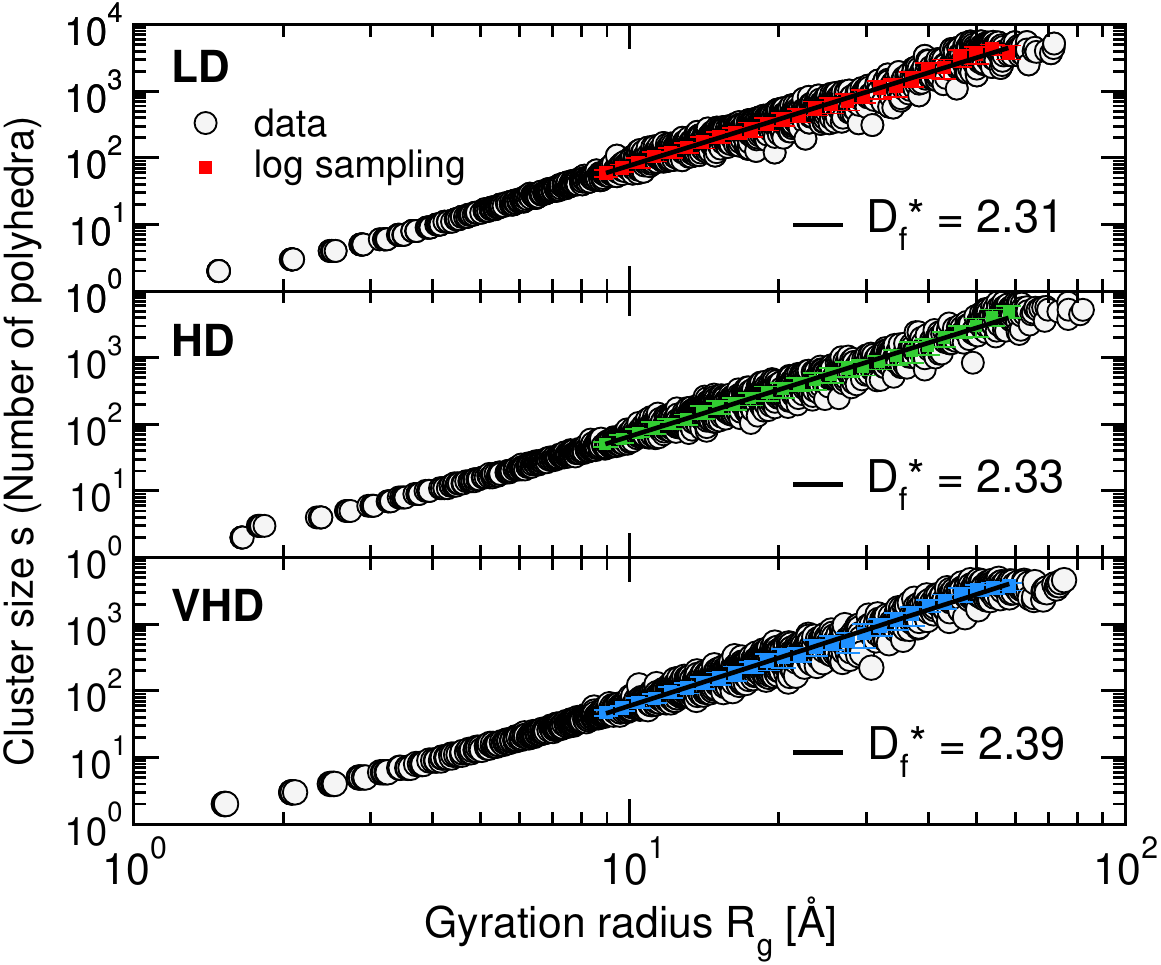}
    \caption{ {\bf Power law in $a$-SiO$_2$:} Gyration radius $R_g$ of non-bonded amorphous silica LD, HD, and VHD clusters at the critical pressure $p_c$ for the system size with 96000 atoms. Lines are fits with a power law $s\propto R_g^{D^*_f}$.}
    \label{fig:SM3}
\end{figure}

\begin{figure}[H]
    \centering
    \includegraphics[width=0.8\linewidth]{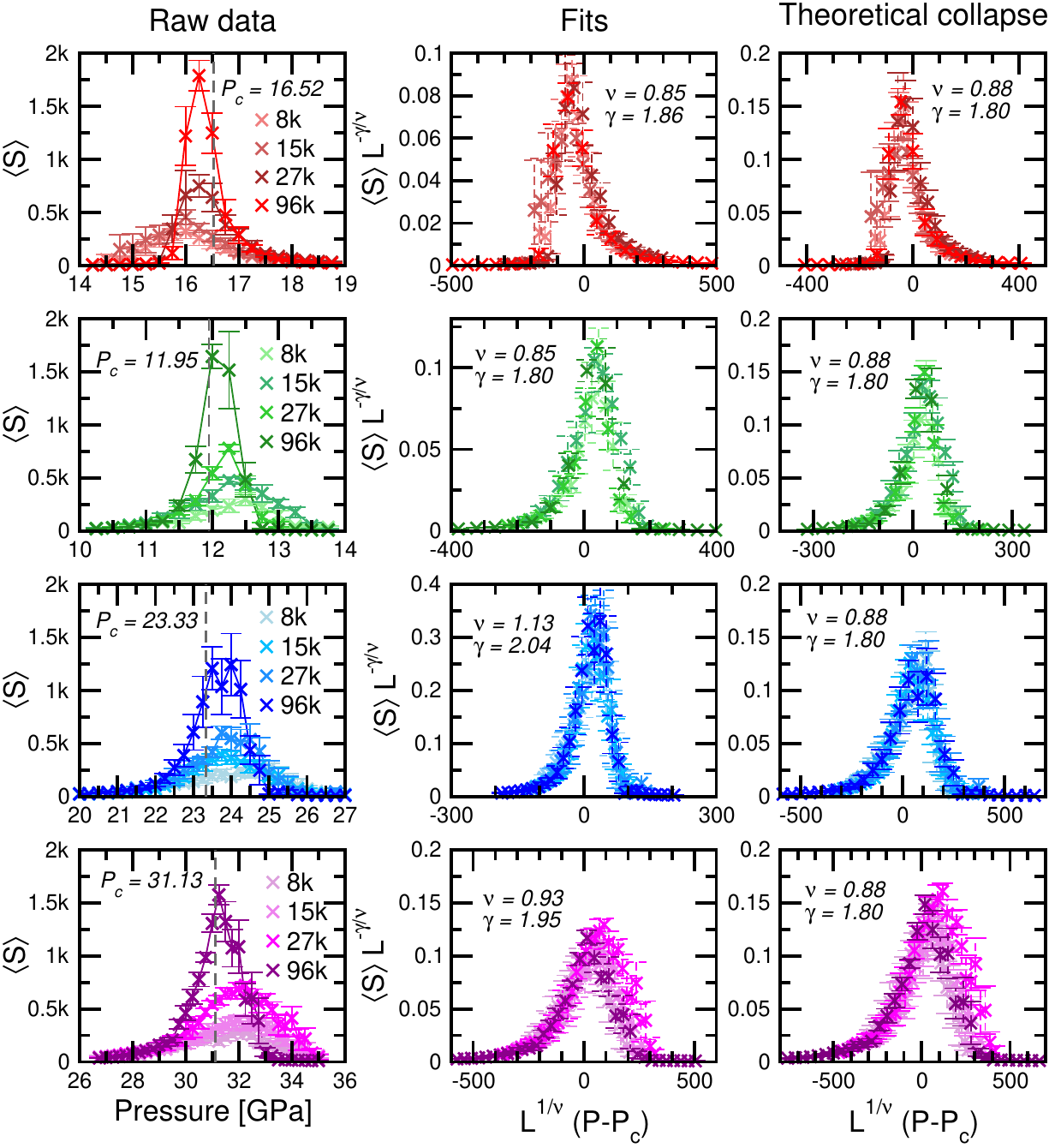}
    \caption{{\bf Data collapse $\langle S\rangle$}. Applying Eq.~(\ref{eq:fssa}) to the raw data (left) yields the single overlapping master curves in the middle. On the right, the collapses with exponents of 3D percolation are displayed for comparison.}
    \label{fig:SM4}
\end{figure}

\ \\
\newpage
\noindent
{\bf Estimation of critical exponents and their uncertainties}\\

The critical exponents and their uncertainties were estimated using two independent procedures. The first consists of fitting the scaling law, with the uncertainties on the data points determined through an error-propagation approach in which two sources of uncertainty were considered and assumed to be independent.
For a given percolation observable $x$ at a given pressure or density, such as the average cluster size $\langle S\rangle$ or the spanning-cluster size $S_{\max}$:
$$
\mathrm{err}(x)
=
\sqrt{
\mathrm{SEM}(x)^2+
\overline{\mathrm{err}(x_i)}^{\,2}
} \quad ,
$$
where $\mathrm{SEM}(x)$ is the variance associated with the mean over the $N=5$ independent samples, 
$$
\mathrm{SEM}(x) =\frac{\sqrt{\frac{1}{N}\sum_i^N(x_i-\overline{x})^2}}{\sqrt{N}}\quad , 
$$
$\mathrm{err}(x_i)$ is the standard error on the mean of $x_i$ over the $M$ configurations within sample $i$ at the given pressure or density. Note that for $a$-SiO$_2$, $M = 51$ for 1k-96k atom samples and $M = 11$ for 1 million atom samples. For $a$-H$_2$O, $M = 10$ for all system sizes.
$$
\mathrm{err}(x_i)=\frac{\sqrt{\frac{1}{M-1}\sum_j^M(x_{i,j}-\overline{x_i})^2}}{\sqrt{M}}\quad ,
$$
and $\overline{\mathrm{err}(x_i)}$ is the mean standard error over the individual samples, 
$$
\overline{\mathrm{err}(x_i)}=\frac{1}{N}\sum_{i=1}^N \mathrm{err}(x_i) \quad .
$$
The critical exponents are obtained by fitting the scaling relation $x(L_k)=f(L_k)$, where $k$ indexes the seven simulation box sizes and each value $x(L_k)$ is associated with its corresponding uncertainty. This procedure does not account for the uncertainty in the critical pressure $p_c$.
\ \\
\ \\
To further assess the robustness of the critical-exponent estimates and their uncertainties, and to account for the uncertainty in the critical pressure $p_c$, we implemented a bootstrap procedure~\cite{Newman_Barkema_1999} based on the five independent samples, using 4,000 resamplings.
At each bootstrap iteration, the five samples available for each system size were resampled with replacement. For each selected sample, the peak position $p_i(L_k)$ and the corresponding peak value $x_i(L_k)$ were resampled together as a pair, $(p_i(L_k),x_i(L_k))$, thereby preserving their correlation. $x_i(L_k)$ refers to the percolation parameters  $\xi$, $\langle S\rangle$, or $S_{\max}$, and $k$ indexes the system sizes.
The peak position, and therefore our estimate of $p_c$, was thus redetermined within each bootstrap iteration. The resulting distribution of the exponents simultaneously incorporates sample-to-sample fluctuations and the uncertainty in the critical point, together with their correlation. The results are summarized in Table SM1, Table SM2, and Table SM3 for the bonded and non-bonded approach (v-SiO$_2$), and non-bonded \textit{a}-H$_2$O, respectively.
\ \\

\begin{table*}[h]
\centering
    \begin{ruledtabular}
    \renewcommand{\arraystretch}{1}
    \begin{tabular}{lcccc}
    Scaling law & \textbf{Error-Propagation} & \textbf{boot-7} & \textbf{boot-6} & \textbf{boot-5} \\
    (connectivity) & \textit{(7 sizes)} & \textit{(7 sizes)} & \textit{(6 sizes)} & \textit{(5 sizes)} \\
    \hline
    \multicolumn{5}{l}{$\xi \propto L^{a}$ \hfill (3D percolation: 1.00)} \\
    \quad SiO$_4$-SiO$_4$   & $0.988 \pm 0.010$ & $0.988 \pm 0.011$ & $0.980 \pm 0.018$ & $0.991 \pm 0.022$ \\
    \quad SiO$_5$-SiO$_5$   & $0.993 \pm 0.020$ & $0.993 \pm 0.016$ & $0.975 \pm 0.022$ & $0.957 \pm 0.028$ \\
    \quad SiO$_6$-SiO$_6$   & $0.946 \pm 0.020$ & $0.946 \pm 0.013$ & $0.937 \pm 0.017$ & $0.938 \pm 0.018$ \\
    \quad Stishovite ($Z=6^*$) & $0.978 \pm 0.010$ & $0.978 \pm 0.010$ & $1.002 \pm 0.019$ & $0.999 \pm 0.019$ \\
    \hline
    \multicolumn{5}{l}{$\langle S\rangle \propto L^{\gamma/\nu}$ \hfill (3D percolation: 2.05)} \\
    \quad SiO$_4$-SiO$_4$   & $2.029 \pm 0.050$ & $2.029 \pm 0.052$ & $2.038 \pm 0.063$ & $2.033 \pm 0.080$ \\
    \quad SiO$_5$-SiO$_5$   & $1.967 \pm 0.050$ & $1.967 \pm 0.043$ & $2.015 \pm 0.075$ & $2.017 \pm 0.075$ \\
    \quad SiO$_6$-SiO$_6$   & $2.001 \pm 0.070$ & $2.001 \pm 0.078$ & $1.982 \pm 0.077$ & $2.004 \pm 0.082$ \\
    \quad Stishovite ($Z=6^*$) & $2.081 \pm 0.040$ & $2.081 \pm 0.038$ & $2.069 \pm 0.037$ & $2.024 \pm 0.044$ \\
    \hline
    \multicolumn{5}{l}{$S_\text{max} \propto L^{D_f}$ \hfill (3D percolation: 2.53)} \\
    \quad SiO$_4$-SiO$_4$   & $2.455 \pm 0.040$ & $2.455 \pm 0.035$ & $2.475 \pm 0.039$ & $2.480 \pm 0.044$ \\
    \quad SiO$_5$-SiO$_5$   & $2.373 \pm 0.040$ & $2.373 \pm 0.047$ & $2.437 \pm 0.048$ & $2.443 \pm 0.046$ \\
    \quad SiO$_6$-SiO$_6$   & $2.398 \pm 0.050$ & $2.398 \pm 0.050$ & $2.401 \pm 0.049$ & $2.438 \pm 0.050$ \\
    \quad Stishovite ($Z=6^*$) & $2.453 \pm 0.030$ & $2.453 \pm 0.029$ & $2.451 \pm 0.029$ & $2.428 \pm 0.029$ \\
    \end{tabular}
    \caption{
    Bonded approach (SiO$_z$-SiO$_z$ structures and stishovite): comparison of
    error-propagation and Bootstrap methods for the finite-size scaling exponents
    and uncertainties.}
    \label{tab:bonded_bootstrap_comparison}
    \end{ruledtabular}
\end{table*}

\begin{table*}[h]
\centering
    \begin{ruledtabular}
    \renewcommand{\arraystretch}{1}
    \begin{tabular}{lcccc}
    Scaling law & \textbf{Error-Propagation} & \textbf{boot-6} & \textbf{boot-5} & \textbf{boot-4} \\
    (density domain) & \textit{(6 sizes)} & \textit{(6 sizes)} & \textit{(5 sizes)} & \textit{(4 sizes)} \\
    \hline
    \multicolumn{5}{l}{$\xi \propto L^{a}$ \hfill (3D percolation: 1.00)} \\
    \quad \textit{LD} & $1.026 \pm 0.026$ & $1.026 \pm 0.026$ & $1.008 \pm 0.036$ & $1.006 \pm 0.035$ \\
    \quad HD          & $0.975 \pm 0.034$ & $0.975 \pm 0.043$ & $0.874 \pm 0.040$ & $0.901 \pm 0.061$ \\
    \quad VHD         & $0.986 \pm 0.025$ & $0.986 \pm 0.028$ & $0.985 \pm 0.033$ & $0.963 \pm 0.025$ \\
    \hline
    \multicolumn{5}{l}{$\langle S\rangle \propto L^{\gamma/\nu}$ \hfill (3D percolation: 2.05)} \\
    \quad \textit{LD} & $2.058 \pm 0.072$ & $2.058 \pm 0.072$ & $2.057 \pm 0.091$ & $1.796 \pm 0.134$ \\
    \quad HD          & $1.923 \pm 0.069$ & $1.923 \pm 0.069$ & $1.822 \pm 0.090$ & $1.799 \pm 0.157$ \\
    \quad VHD         & $2.053 \pm 0.042$ & $2.053 \pm 0.035$ & $2.016 \pm 0.050$ & $1.999 \pm 0.074$ \\
    \hline
    \multicolumn{5}{l}{$S_\text{max} \propto L^{D_f}$ \hfill (3D percolation: 2.53)} \\
    \quad \textit{LD} & $2.393 \pm 0.055$ & $2.393 \pm 0.054$ & $2.422 \pm 0.075$ & $2.255 \pm 0.105$ \\
    \quad HD          & $2.297 \pm 0.059$ & $2.297 \pm 0.059$ & $2.243 \pm 0.073$ & $2.196 \pm 0.120$ \\
    \quad VHD         & $2.387 \pm 0.037$ & $2.387 \pm 0.034$ & $2.388 \pm 0.048$ & $2.387 \pm 0.057$ \\
    \end{tabular}
    \caption{
    Non-bonded approach (SiSi$_z$-SiSi$_z$ density domains): comparison of
    error-propagation and Bootstrap methods for the finite-size scaling exponents
    and uncertainties.}
    \label{tab:non_bonded_bootstrap_comparison}
    \end{ruledtabular}
\end{table*}

\begin{figure}[H]
\centering
\includegraphics[width=15cm]{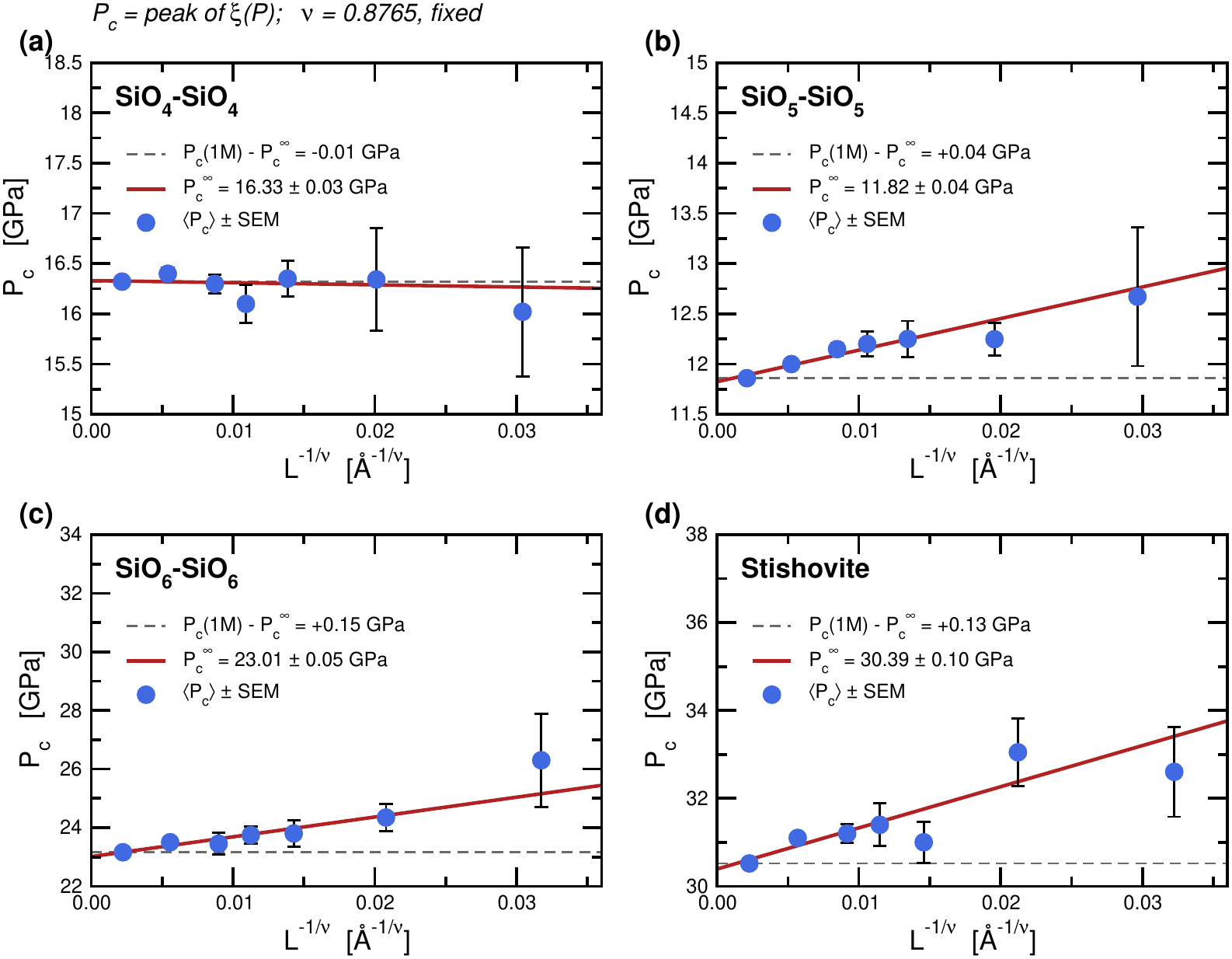}
\caption{$p_c$ versus $L^{-1/\nu}$. Mean of the peak-based $p_c$ values derived from the maximum of $\xi$ (blue circles) and fitted using Eq.~(\ref{eq:pc_extrap}) (red line). The intercept at the origin yields $p_c^\infty$.}
\label{fig:SM5}
\end{figure}

\begin{table}[h]
\centering
    \begin{ruledtabular}
    \renewcommand{\arraystretch}{1}
    \begin{tabular}{lcccc}
    Scaling law & \textbf{Error-Propagation} & \textbf{boot-5} & \textbf{boot-4} & \textbf{boot-3} \\
    (density domain) & \textit{(5 sizes)} & \textit{(5 sizes)} & \textit{(4 sizes)} & \textit{(3 sizes)} \\
    \hline
    \multicolumn{5}{l}{$\xi \propto L^{a}$ \hfill (3D percolation: 1.00)} \\
    \quad \textit{LD}      & $0.969 \pm 0.066$    & $0.969 \pm 0.048$    & $0.998 \pm 0.064$    & $0.801 \pm 0.127$    \\
    \quad HD               & $1.051 \pm 0.036$    & $1.051 \pm 0.024$    & $1.018 \pm 0.029$    & $0.946 \pm 0.063$    \\
    \quad VHD              & $0.997 \pm 0.042$    & $0.997 \pm 0.023$    & $0.937 \pm 0.036$    & $0.891 \pm 0.050$    \\
    \hline
    \multicolumn{5}{l}{$\langle S\rangle \propto L^{\gamma/\nu}$ \hfill (3D percolation: 2.05)} \\
    \quad \textit{LD}      & $2.090 \pm 0.138$    & $2.090 \pm 0.092$    & $2.190 \pm 0.087$    & $1.374 \pm 0.201$    \\
    \quad HD               & $2.079 \pm 0.089$    & $2.079 \pm 0.055$    & $2.067 \pm 0.113$    & $2.191 \pm 0.217$    \\
    \quad VHD              & $2.070 \pm 0.108$    & $2.070 \pm 0.070$    & $2.056 \pm 0.099$    & $1.830 \pm 0.155$    \\
    \hline
    \multicolumn{5}{l}{$S_\text{max} \propto L^{D_f}$ \hfill (3D percolation: 2.53)} \\
    \quad \textit{LD}      & $2.454 \pm 0.116$    & $2.454 \pm 0.077$    & $2.568 \pm 0.069$    & $1.928 \pm 0.169$    \\
    \quad HD               & $2.384 \pm 0.072$    & $2.384 \pm 0.050$    & $2.386 \pm 0.084$    & $2.447 \pm 0.179$    \\
    \quad VHD              & $2.312 \pm 0.086$    & $2.312 \pm 0.059$    & $2.327 \pm 0.085$    & $2.213 \pm 0.124$    \\
    \end{tabular}
    \caption{
    Amorphous water at 124\,K (HD / LD / VHD density domains): comparison of
    error-propagation and Bootstrap methods for the finite-size scaling exponents
    and uncertainties.}
    \label{tab:aH2O_bootstrap_comparison}
    \end{ruledtabular}
\end{table}
\newpage
\noindent
{\bf Bonded SiO$_2$ under compression: influence of the critical pressure $p_c$ on the fractal dimension $D_f$}\\
\ \\
\indent
The fractal dimension $D_f$ of the spanning cluster is highly sensitive to the value of $p_c$, so it is important to assess how its estimate varies with the location of $p_c$. 
For this purpose, we first determined, for each percolation transition, the asymptotic critical pressure in the infinite-system limit, $p_c^\infty$. For each sample, $p_c(L)$ was defined as the pressure corresponding to the maximum of the quantity used in the finite-size scaling analysis, namely $\xi(L)$, $\langle S\rangle(L)$, or $S_{\max}(L)$.
For each system size, the five independent samples were averaged, yielding the {\it peak-based} estimate of $p_c(L)$ used throughout the article. Given that the condition $\xi = 1$ is fairly well satisfied for all percolation transitions, the resulting finite-size dependence of $p_c(L)$ was then fitted using 
\begin{equation}
p_c(L) = p_c^\infty + A\,L^{-1/\nu},
\qquad
\nu = 0.8765 \ \text{(fixed)},
\label{eq:pc_extrap}
\end{equation}
with system sizes ranging from 1008 to 1\,056\,000 atoms.\\
\noindent
In Fig.~\ref{fig:SM5}, the plots of $p_c$ extracted from the peak position of the correlation length $\xi$ clearly show an increasing overestimation of $p_c$ as the box size decreases, at least for SiO$_5$, SiO$_6$ and stishovite spanning clusters. Extrapolation to $L^{-1/\nu} \rightarrow 0$ using Eq.~(\ref{eq:pc_extrap}) yields the asymptotic critical pressure $p_c^\infty$.  
It is worth noting that, since $\nu$ is fixed at the standard percolation value, $\nu = 0.8765$, the resulting values of $p_c^\infty$ are not entirely independent of the percolation assumption. 
However, the hyperscaling closure discussed below makes this a genuine consistency test rather than a mere assumption.
\noindent
Table\,SM4 shows that, within the proposed approach, the peak-based estimates of $p_c$ for the 1\,056\,000-atom system closely approach the extrapolated values $p_c^\infty$.
\begin{table}[H]
\centering
\begin{tabular}{l cc}
\toprule
 & peak-based $p_c(\mathrm{1M})$ & $p_c^\infty$ \\
\midrule
SiO$_4$--SiO$_4$ & $16.32\pm0.02$ & $16.331\pm0.035$ \\
SiO$_5$--SiO$_5$ & $11.86\pm0.04$ & $11.824\pm0.043$ \\
SiO$_6$--SiO$_6$ & $23.16\pm0.03$ & $23.011\pm0.048$ \\
Stishovite & $30.52\pm0.09$ & $30.393\pm0.105$ \\
\bottomrule
\end{tabular}
\caption{Peak-based $p_c$ of the 1\,056\,000-atom system (mean $\pm$ SEM) and extrapolated $p_c^\infty$, in GPa.}
\end{table}
\noindent
Table\ SM5 compares the  $D_f$  resulting from the scaling law of the spanning cluster size ($S_{max}\propto L^{D_f}$) and the average cluster size $\langle S \rangle$, respectively, using $p_c$ obtained by the peak position for each transition (Table II in the article) and $p_c$ fixed to $p_c^\infty$.  
The results first show that a more accurate estimate of $p_c$ substantially improves the agreement of $D_f$ with the standard percolation value.
In addition, Table\,SM6 compares with the value of $D_f$ derived from the hyperscaling relation, $D_f=(3+\gamma/\nu)/2$, showing that the deviations in the exponents obtained using the peak-based estimate of $p_c$ are consistent in direction with the corresponding hyperscaling mismatch.
The agreement between the values of $D_f$ obtained from hyperscaling and from finite-size scaling demonstrates the internal consistency of the analysis and supports the interpretation of a one-sided bias in the peak-based estimate of $p_c$, rather than evidence for a distinct universality class.
\begin{table}[H] 
\centering  
\small 
\begin{tabular}{l cc} 
\toprule
& peak-based $p_c$ & $p_c^\infty$ \\ 
\midrule 
SiO$_4$--SiO$_4$ & $2.455\pm0.035$ & $2.584\pm0.068$ \\ SiO$_5$--SiO$_5$ & $2.373\pm0.047$ & $2.483\pm0.115$ \\ SiO$_6$--SiO$_6$ & $2.398\pm0.050$ & $2.444\pm0.133$ \\ Stishovite & $2.453\pm0.029$ & $2.505\pm0.143$ \\ \bottomrule 
\end{tabular} 
\caption{$D_f$ derived from the scaling relation, $S_{\max} \sim L^{D_f}$, using  the peak-based $p_c(L)$ estimates and the extrapolated $p_c^\infty$ values (standard percolation $D_f=2.53$).}
\end{table}
\begin{table}[H]
\centering
\begin{tabular}{l ccc ccc}
\toprule
 & \multicolumn{3}{c}{peak-based $p_c$} & \multicolumn{3}{c}{$p_c^\infty$} \\
\cmidrule(lr){2-4}\cmidrule(lr){5-7}
 & $\gamma/\nu$ & $(3+\gamma/\nu)/2$ & $D_f$ & $\gamma/\nu$ & $(3+\gamma/\nu)/2$ & $D_f$ \\
\midrule
SiO$_4$--SiO$_4$ & $2.029 \pm 0.052$ & 2.514 & 2.455 & $2.172 \pm 0.112$ & 2.586 & 2.584 \\
SiO$_5$--SiO$_5$ & $1.967 \pm 0.043$ & 2.484 & 2.373 & $2.001 \pm 0.185$ & 2.500 & 2.483 \\
SiO$_6$--SiO$_6$ & $2.001 \pm 0.078$ & 2.500 & 2.398 & $1.934 \pm 0.198$ & 2.467 & 2.444 \\
Stishovite & $2.081 \pm 0.038$ & 2.540 & 2.453 & $2.006 \pm 0.244$ & 2.503 & 2.505 \\
\bottomrule
\end{tabular}
\caption{Hyperscaling: comparison of $D_f$ obtained from the peak-based and $p_c^\infty$ analyses with the values derived from the hyperscaling relation, $D_f=(d+\gamma/\nu)/2$, with $d=3$.}
\end{table}

\end{document}